\documentclass[reprint, superscriptaddress, nofootinbib, amsmath, amssymb, aps, prl, floatfix]{revtex4-2}
\usepackage{graphicx} 
\usepackage{dcolumn}
\usepackage{bm} 
\usepackage{braket}
\usepackage{dsfont}
\usepackage{color,colortbl}
\usepackage[table,xcdraw]{xcolor}
\usepackage{hyperref}
\hypersetup{
colorlinks=true,
urlcolor= blue,
citecolor=blue,
linkcolor= blue,
}
\usepackage{multirow}
\usepackage{mwe}
\usepackage{booktabs}
\usepackage{caption}
\usepackage{lipsum}
\usepackage{babel,blindtext}
\usepackage[toc,page]{appendix}
\usepackage{amsmath}
\usepackage[version=4]{mhchem}
\usepackage{physics}
\usepackage{array}
\usepackage{placeins}

\begin{document}

\title{Revealing the proton slingshot mechanism in solid acid electrolytes through machine learning molecular dynamics} 

\author{Menghang Wang}
\affiliation{Harvard John A. Paulson School of Engineering and Applied Sciences, Harvard University, Cambridge, MA 02138, USA}

\author{Jingxuan Ding}
\affiliation{Harvard John A. Paulson School of Engineering and Applied Sciences, Harvard University, Cambridge, MA 02138, USA}

\author{Grace Xiong}
\affiliation{Materials Science and Engineering, Northwestern University, Evanston, IL 60208, USA}

\author{Ni Zhan}
\affiliation{Department of Computer Science, Princeton University, Princeton, NJ 08544, USA}

\author{Cameron J. Owen}
\affiliation{Department of Chemistry and Chemical Biology, Harvard University, Cambridge, MA 02138, USA}

\author{Albert Musaelian}
\affiliation{Harvard John A. Paulson School of Engineering and Applied Sciences, Harvard University, Cambridge, MA 02138, USA}

\author{Yu Xie}
\affiliation{Harvard John A. Paulson School of Engineering and Applied Sciences, Harvard University, Cambridge, MA 02138, USA}

\author{Nicola Molinari}
\affiliation{Harvard John A. Paulson School of Engineering and Applied Sciences, Harvard University, Cambridge, MA 02138, USA}
\affiliation{Robert Bosch Research and Technology Center, Watertown, MA 02472, USA}

\author{Ryan P. Adams}
\affiliation{Department of Computer Science, Princeton University, Princeton, NJ 08544, USA}

\author{Sossina Haile}
\affiliation{Materials Science and Engineering, Northwestern University, Evanston, IL 60208, USA}

\author{Boris Kozinsky}
\affiliation{Harvard John A. Paulson School of Engineering and Applied Sciences, Harvard University, Cambridge, MA 02138, USA}
\affiliation{Robert Bosch Research and Technology Center, Watertown, MA 02472, USA}

\begin{abstract}
In solid acid electrolytes \ce{CsH2PO4} and \ce{CsHSO4}, the origin of fast proton conduction has long been debated, attributed either to local proton hopping or to polyanion rotation. However, the precise role of polyanion rotation and its interplay with proton hopping remained unresolved. Nanosecond-scale molecular dynamics simulations, driven by equivariant neural network force fields, reveal a nuanced proton slingshot mechanism: protons are first carried by rotating polyanions, followed by \ce{O-H} bond reorientation, and the combined motion enables long-range jumps. This mechanism challenges the conventional revolving paddlewheel model, revealing long-range transport arises from significant \ce{O-H} bond reorientation assisted by only limited polyanion rotation. Despite their structural similarity, \ce{CsH2PO4} and \ce{CsHSO4} exhibit qualitatively different transport behavior, governed by their distinct proton concentrations. In \ce{CsH2PO4}, two rotation rates with different activation energies emerge, while \ce{CsHSO4} displays a single-rate process with faster productive polyanion dynamics. The higher proton concentration in \ce{CsH2PO4} frustrates \ce{PO4} rotation, resulting in slower dynamics than for \ce{SO4} in \ce{CsHSO4}. In \ce{CsH2PO4}, we further identify a unique correlation between oxygen sharing and proton transport, arising from the extra proton per polyanion. Together, our findings establish a unified framework linking polyanion rotation, proton coordination, and long-range transport, and suggest that tuning proton concentration could accelerate rotation and enhance conductivity. 
\end{abstract}

\maketitle

{\renewcommand{\thefootnote}{}%
\footnotetext{This document is the unedited Author's version of a Submitted Manuscript subsequently accepted for publication in Journal of the American Chemical Society, copyright \copyright{} 2026 American Chemical Society. To access the final published article, see \url{https://doi.org/10.1021/jacs.5c21965}. This version was submitted on 17 December 2025, before peer review, and does not include the revisions. The Data and Code availability statements, however, are up to date.}}

\section{Introduction}

Several solid acid compounds exhibit high proton conductivity as a result of structural disorder \cite{Baranov_2003}. In addition to high ionic conductivity, these compounds typically display negligible electronic conductivity, anhydrous proton transport, and impermeability to fuels, rendering solid acids attractive as electrolytes in a range of electrochemical energy devices \cite{Haile_2001, Boysen_2004}. Technological development efforts have focused on \ce{CsH2PO4} (cesium dihydrogen phosphate, CDP) due to its stability under both oxidizing and reducing conditions \cite{Boysen_2004, Merle_2002}. The material has been demonstrated in pre-commercial fuel cell stacks \cite{Gittleman_2021}, in hydrogen electrochemical pumps \cite{Papandrew2014}, and in ammonia electrolysis cells for on-site production of high purity hydrogen \cite{Lim_2020}. 

\begin{figure*}[htbp!]
\centering
\includegraphics[width=0.7\linewidth]{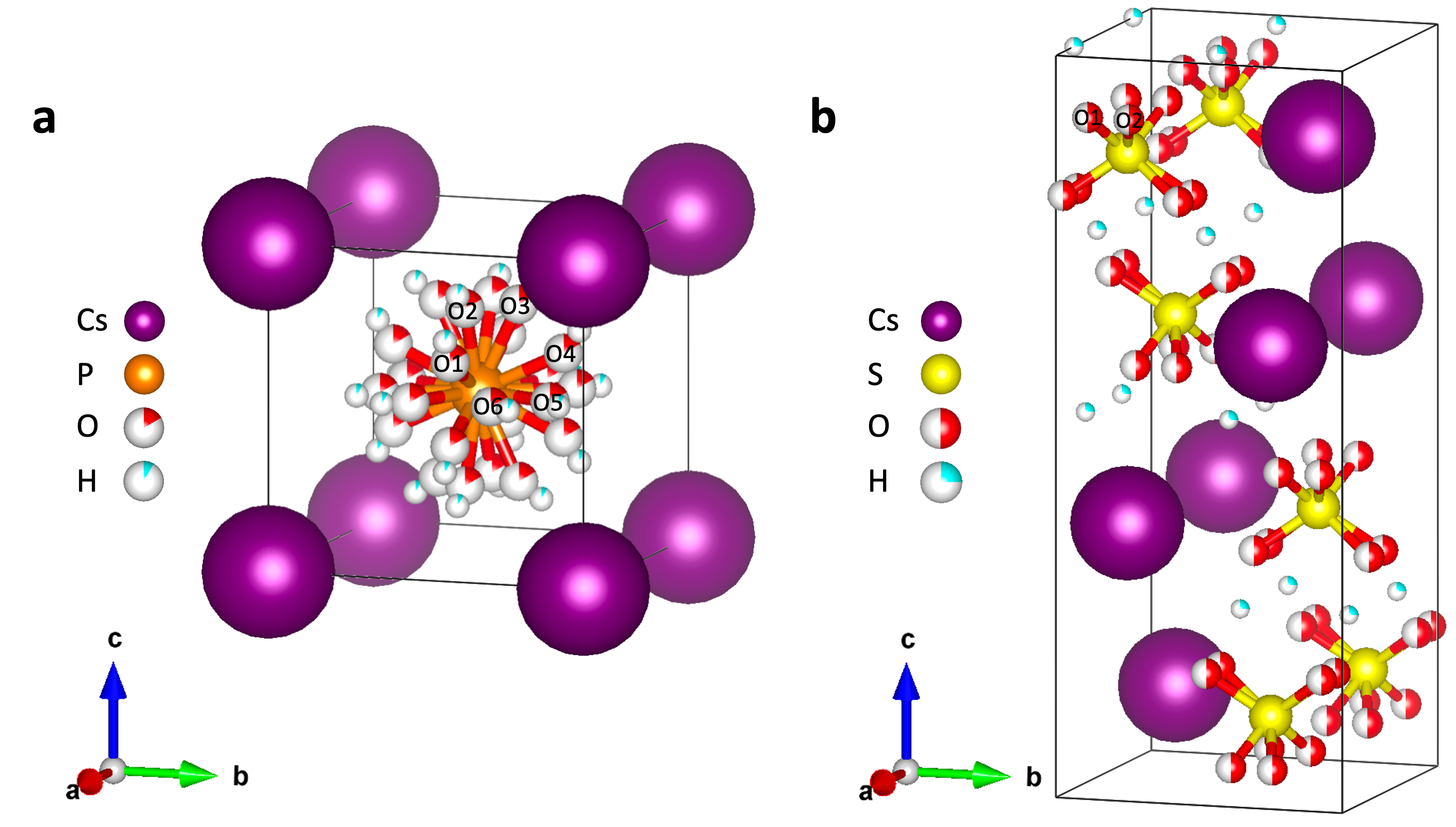}
\caption{\textbf{Superprotonic structures.} The superprotonic phase of \textbf{(a)} \ce{CsH2PO4} \cite{Yamada_2004} and \textbf{(b)} \ce{CsHSO4} \cite{Jirak_1987}. The oxygen sites in \ce{CsH2PO4} have a one-sixth average occupancy, whereas those in \ce{CsHSO4} have a one-half average occupancy. Both \ce{PO4} and \ce{SO4} polyanions are orientationally disordered. The \ce{H} position in \ce{CsH2PO4} is determined according to \cite{Lee2008}, with details provided in Supporting Information.}
\label{fig::struct}
\end{figure*}

The essential structural features that result in high conductivity in \ce{CsH2PO4} and related compounds such as \ce{CsHSO4} (cesium hydrogen sulfate, CHS) have been under investigation for several decades, ever since the discovery of the spectacular (several orders of magnitude) increase in conductivity at the transition to the disordered state \cite{Baranov1982, Baranov1988}. In a general sense, proton transport is known to occur via a mechanism involving a combination of local proton hops along \ce{O-H\bond{...}O} bonds and rotation of the \ce{XO4} polyanion group (X = P, S, As, Se, etc.), a Grotthuss-type mechanism \cite{Kreuer_1996, Norby_1999}. In the high temperature, superprotonic structures, both steps become rapid, irrespective of the structural details. In particular, \ce{CsH2PO4} adopts a CsCl structure, characterized by cubic (Pm$\bar{3}$m) symmetry \cite{Yamada_2004, Botez_2007, Bronowska1990}, whereas \ce{CsHSO4} adopts a Scheelite structure, characterized by tetragonal (I$4_1$/amd) symmetry \cite{Jirak_1987}. In both materials (Figure \ref{fig::struct}a,b), the number of crystallographically equivalent oxygen sites exceeds the number of oxygen atoms available to fill them, supporting orientational disorder in the \ce{PO4} and \ce{SO4} polyanion groups \cite{Haile_2007}. 

The occurrence of polyanion rotation in these structures has been directly observed by nuclear magnetic resonance (NMR) spectroscopy \cite{Ishikawa2008, Yamada_2004, Blinc_1995}, quasielastic neutron scattering (QENS) \cite{Ishikawa2008, Belushkin_1992}, measurements of dielectric relaxation \cite{Badot_1989}, and MD studies \cite{Chisholm_2005, Wood2007, Lee2008, Dressler2020, Jinnouchi2022}. Therefore, polyanion rotation has long been recognized as the key contributor to fast proton mobility.

MD studies have suggested that the rate limiting mechanisms for proton motion in superprotonic \ce{CsH2PO4} and \ce{CsHSO4} differ \cite{Lee2008, Wood2007, Dressler2020}. Specifically, polyanion rotation has been hypothesized as rate-limiting in \ce{CsH2PO4}, while local proton hopping is rate-limiting in \ce{CsHSO4} \cite{Dressler2020}. However, these steps have generally been considered independently, and the possible coupling between polyanion rotation and proton motion has not been explored. In particular, it remains unclear whether large-angle rotations are necessary for efficient proton transport or if small-scale rotations are sufficient. Conventional \textit{ab initio} molecular dynamics (AIMD) simulations have been unable to resolve this, as the absence of observed large-amplitude motions likely stems from limited timescales and system sizes rather than their physical absence \cite{Lee2008, Wood2007}.

The reason for differing rate-limiting steps in \ce{CsH2PO4} and \ce{CsHSO4} is also unclear. Besides the symmetry differences in the superprotonic phases, these compounds are distinct in terms of proton concentration. The stoichiometric differences imply that the number of hydrogen bonds per polyanion group in \ce{CsH2PO4} (4) is double that of \ce{CsHSO4} (2). Concomitantly, in \ce{CsH2PO4}, the average number of covalently bonded \ce{O-H} bonds per polyanion group is 2, whereas it is only 1 in \ce{CsHSO4}. Whether this is the reason for faster polyanion reorientation in \ce{CsHSO4} is unknown. The proton excess compound \ce{Cs7(H4PO4)(H2PO4)8} adopts cubic symmetry and a structural arrangement similar to that of \ce{CsH2PO4} \cite{Wang_2020}. In this material, significant differences in rotation rates are observed between \ce{H4PO4+} polycations and \ce{H2PO4-} polyanions \cite{Dressler_2023}, highlighting the significant impact of local proton coordination on anion dynamics.

Direct investigation of the interplay between polyanion rotation and proton transport faces significant computational challenges. Conventional AIMD simulations are generally limited to small system sizes (up to 200 atoms) and short durations (up to 250 ps), restricting their ability to capture long-timescale or collective effects \cite{Dressler2020}. Accurate modeling of polyanion orientational disorder in the superprotonic phases demands larger simulation cells to minimize finite size effects \cite{Lee2008}. Moreover, achieving sufficient sampling of proton diffusion events in AIMD often necessitates elevated temperatures that exceed the thermal stability of these materials.

Machine learning force fields (MLFFs) significantly expand the reach of MD simulations while preserving \textit{ab initio} accuracy \cite{Behler2007, Vandermause_2022, Xie_2023, Batzner_2022, Musaelian2023}. By learning the mapping from atomic geometries to density functional theory (DFT) forces, MLFFs replace costly DFT evaluations at each MD step, enabling simulations over longer timescales and larger system sizes than conventional AIMD. Recent applications to proton-conducting media containing phosphoric acid \cite{Jinnouchi2022, Minami2023} have revealed the intricate interplay between polyanion rotation and proton bonding, with Minami \textit{et al.} demonstrating that hydrogen bond recombination promotes polyanion rotations \cite{Minami2023}. These results indicate that proton transport cannot be fully captured by the traditional two-step hopping-rotation picture.

In this work, we leverage our recently developed MLFF techniques to quantitatively resolve proton conduction mechanisms in the superprotonic phases of \ce{CsH2PO4} and \ce{CsHSO4}. Through nanosecond simulations of thousands of atoms at \textit{ab initio} precision, we uncover a novel ``slingshot" transport mechanism, in which synchronized polyanion rotation and \ce{O-H} bond reorientation drive long-range proton diffusion. In \ce{CsH2PO4}, we identify two distinct polyanion rotation rates with different activation energies, a behavior correlated with fluctuations in local proton coordination. This material is further characterized by rotational frustration. In contrast, \ce{CsHSO4} exhibits simpler dynamics: the \ce{SO4} polyanions adopt distinct, preferred orientations and rotate at a single characteristic rate. By uncovering the coupled dynamics of polyanion rotation, \ce{O-H} bond reorientation, and proton coordination, our findings establish a unified mechanistic framework and provide design principles for next-generation solid-state electrolytes with enhanced performance.

\section{Method} \label{sec:method}

\begin{figure*}[htbp!]
\centering
\includegraphics[width=\linewidth]{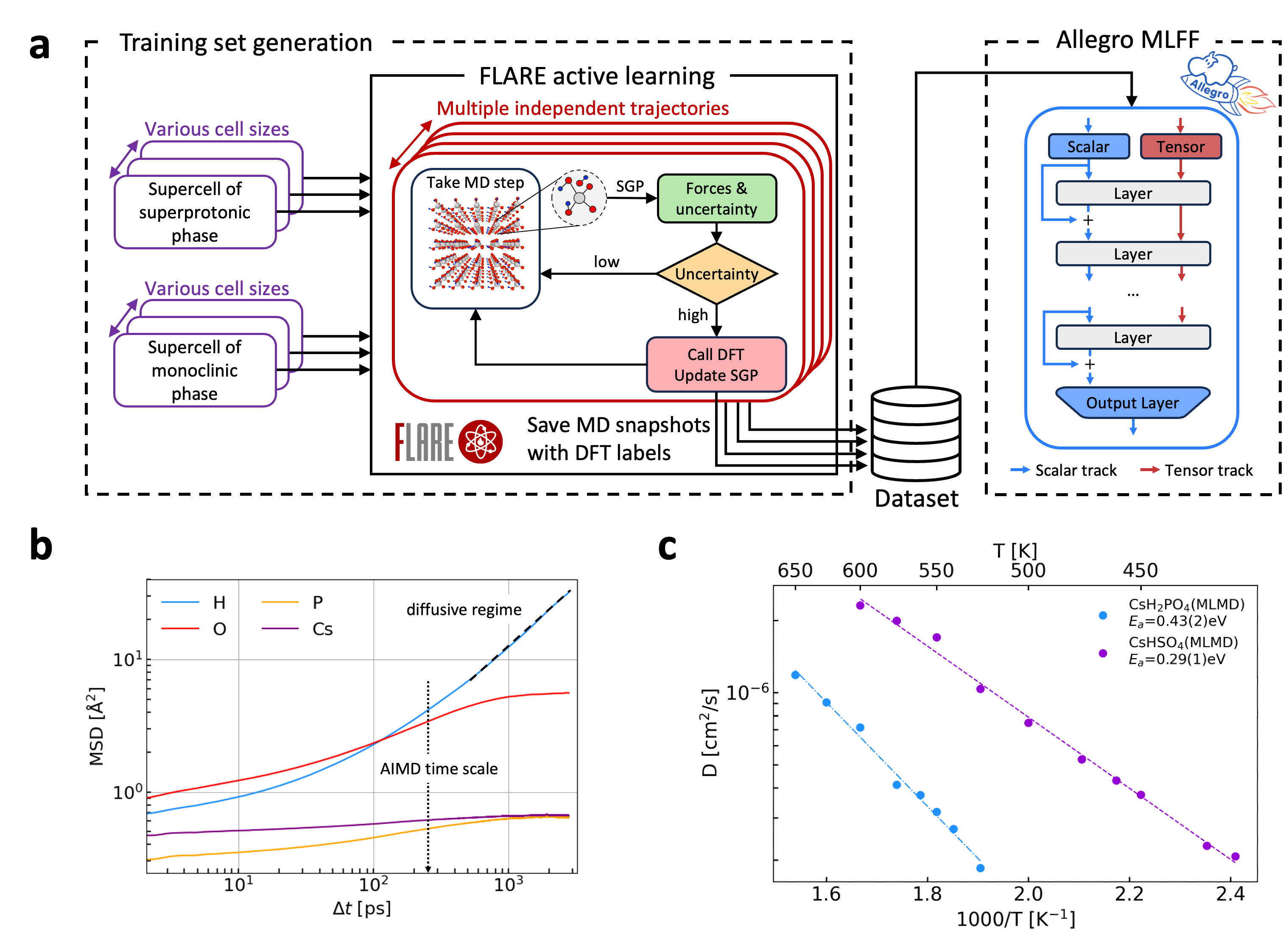}
\caption{\textbf{MLFF workflow, diffusion coefficients, and MSD.} \textbf{(a)} Left: workflow overview for generating diverse training data using FLARE Bayesian active learning \cite{Vandermause_2020, Xie_2023}. Right: Allegro equivariant MLFF architecture \cite{Musaelian2023}. \textbf{(b)} MSD of each species in \ce{CsH2PO4} from a 4 ns ML MD at 525 K. Typical AIMD simulations have not yet reached the proton diffusive regime (dashed line). \textbf{(c)} For \ce{CsH2PO4} (CDP), the proton diffusion activation energy from ML MD is 0.43(2) eV, consistent with experimental values of 0.39 \cite{Ishikawa2008}, 0.398 \cite{Ikeda_2014}, and 0.43 eV \cite{Haile_2007}. For \ce{CsHSO4} (CHS), the activation energy from ML MD is 0.29(1) eV, also in agreement with experimental values of 0.29 \cite{Blinc_1995}, 0.33 \cite{Baranov1982}, and 0.36 eV \cite{HAYASHI_2004}.}
\label{fig::method_Ea_msd}
\end{figure*}

Our investigation is made possible by an accurate and reliable MLFF capable of capturing ion dynamics over long timescales. We trained the equivariant MLFF Allegro \cite{Musaelian2023} on a diverse dataset of structures with corresponding DFT energies and forces, collected through the Bayesian active learning framework FLARE \cite{Vandermause_2020, Vandermause_2022, Xie_2023}. 

DFT calculations were performed with the Perdew-Burke-Ernzerhof (PBE) \cite{Perdew_1996} functional, which accurately describes material properties in both monoclinic phases \cite{Troeye_2017, Krzystyniak_2015} and superprotonic phases \cite{Parker_2020, Lee2008, Sassnick2021, Wood2007, Dressler2020} of \ce{CsH2PO4} and \ce{CsHSO4}. Calculations were executed using the Vienna \textit{ab initio} simulation package (VASP) \cite{Kresse1996_1, Kresse1996_2, Kresse1999}, with detailed DFT parameters provided in Supporting Information.

Diverse training data are essential for stable MLFFs. While short AIMD trajectories often fail to capture varied polyanion orientations and rare proton jump events, FLARE automates data collection and ensures data diversity. FLARE employs rotationally invariant B2 descriptors from the atomic cluster expansion (ACE) \cite{Drautz_2019} to describe atomic environments and uses sparse Gaussian processes (SGP) \cite{Vandermause_2022} to construct a Bayesian force field (BFF). The BFF predicts atomic forces and uncertainties on local atomic environments during an active learning MD simulation. When uncertainties exceed a set threshold, FLARE invokes DFT to calculate interatomic forces, energy, and stresses for the MD snapshot and updates both the training set and the SGP model.

The dataset generation, shown in Figure \ref{fig::method_Ea_msd}a, consists of multiple active learning trajectories covering varying temperatures, pressures, and supercell sizes for superprotonic and monoclinic phases, each initiated with an empty SGP model. The dataset of each compound was split into training (80\%), validation (10\%), and test (10\%) sets, maintaining proportional representation across different active learning trajectories. This dataset forms the foundation for training the Allegro MLFF. Further details on dataset generation are described in Supporting Information.

Neural network-based MLFFs often yield higher accuracy than models based on hand-crafted descriptors, though challenges in efficiency and robustness persist. Equivariant MLFFs \cite{Batzner_2022, Musaelian2023}, such as Allegro \cite{Musaelian2023}, have shown low error, high sample efficiency, and robustness across various materials. Allegro combines symmetry-invariant and equivariant features through tensor products, depicted in Figure \ref{fig::method_Ea_msd}a, in a deep equivariant graph neural network, achieving superior accuracy and reliability compared to models using only invariant information \cite{Fu2022}. 

We trained two Allegro MLFFs using identical architectures, one on the \ce{CsH2PO4} dataset and one on the \ce{CsHSO4} dataset. Each model consists of 2 layers with 32 features under irreducible representations of $O(3)$ symmetry group and a $l_{\text{max}}$ of 2. The 2-body latent multi-layer perceptron (MLP) consists of three layers of dimension [32, 64, 128], using SiLU nonlinearities in the hidden layers. The latter latent MLPs consist of two hidden layers of dimension [128,128]. The final MLP has the dimension of 32 and no nonlinearity. MLPs were initialized with a uniform distribution of unit variance. We applied a radial cutoff of 7.0 Å for local atomic environments. 

Each model was trained using a joint loss function combining the mean square error of energies and forces:
\begin{equation}
\mathcal{L} = \frac{\lambda_E}{B} \sum_b^B \left( \hat{E}_b - E_b \right)^2 + \frac{\lambda_F}{3BN} \sum_i^{BN} \sum_{\alpha=1}^3 \|-\frac{\partial \hat{E} }{\partial r_{i,\alpha}}- F_{i,\alpha}\|^2
\end{equation}
where $B$ is the batch size, $N$ is the number of atoms, $E_{b},\hat{E}_b$ are the true and predicted energies of a given batch, and $F_{i,\alpha}$ is the force component on atom $i$ in direction $\alpha$. Energy and force losses were weighted equally with $\lambda_E=1$ and $\lambda_F=1$. We applied a learning rate of 0.002 and a batch size of 5, using the Adam optimizer \cite{adam} in PyTorch \cite{pytorch}, with default parameters ($\beta_1$ = 0.9, $\beta_2$ = 0.999, $\epsilon=10^{-8}$, no weight decay). 

We used an on-plateau scheduler, with a patience of 50 and a decay factor of 0.8, reducing the learning rate based on the validation loss. An exponential moving average with a weight of 0.99 was used for the final model and for evaluation on the validation set. All models were trained with float32 precision on a single NVIDIA A100 GPU.

The mean absolute errors on test sets are 0.64 and 1.6 meV/atom in energy, and 31 and 56 meV/Å in forces for \ce{CsH2PO4} and \ce{CsHSO4}, respectively. The force parity plots (Figure S1, Supporting Information) demonstrate excellent agreements between Allegro predictions and DFT-calculated forces. To evaluate robustness, we varied cell parameters in the superprotonic phases to predict the equation of state for both compounds. Even for cell volumes outside the training set, Allegro achieved predictive errors within 2 meV/atom (Figure S2, Supporting Information).

We use the trained Allegro MLFF to drive MD simulations of the superprotonic phase in the \textit{NVT} ensemble with LAMMPS \cite{Thompson_2022}. Simulation cells contain 1,000 atoms for \ce{CsH2PO4} and 1,344 atoms for \ce{CsHSO4}, with trajectories of 3--4 ns at a 0.5 fs timestep. We save ML MD snapshots every 50 fs and equilibrate the system for 10 ps before analysis. Experimental unit cell parameters were employed (4.96 Å for \ce{CsH2PO4} in Pm$\bar{3}$m symmetry \cite{Yamada_2004}, $a = b = 5.72$ Å, $c = 14.23$ Å for \ce{CsHSO4} in I$4_1$/amd symmetry \cite{Jirak_1987}), and supercells were constructed by duplication. Additional details are provided in the Supporting Information.

\section{Results} 

The methodological framework described above enables nanosecond-scale MD simulations essential for investigating proton transport mechanisms. Here, we first analyze diffusion characteristics and activation energies of both compounds.

\subsection{Diffusion coefficients and activation energies}

Figure \ref{fig::method_Ea_msd}b illustrates that in \ce{CsH2PO4}, protons require approximately 500 ps to reach a fully diffusive regime, where the mean square displacement (MSD) follows Fickian diffusion ($\alpha=1$ for MSD(H) $\sim (\Delta t)^\alpha$). In contrast, the oxygen MSD plateaus only after 1 ns, indicating that simulations must extend to nanosecond timescales to adequately sample the range of polyanion configurations. Similarly, for \ce{CsHSO4}, the MSD analysis (Figure S5, Supporting Information) shows that proton diffusion emerges only above 200 ps. These findings suggest that most previous AIMD studies, which are limited to shorter timescales, probe only the initial, sub-diffusive motion of protons and are therefore unable to fully resolve the relevant transport mechanisms or reliably predict proton diffusivity.

We validate our ML MD simulations by comparing computed proton diffusion coefficients and Arrhenius activation energies with experimental measurements. Our ML MD-derived diffusion coefficients show significantly better agreement with experiments compared to previous AIMD studies, which deviated by an order of magnitude \cite{Dressler2020, Wood2007, Lee2008}. The calculated activation energies of 0.43(2) eV for \ce{CsH2PO4} and 0.29(1) eV for \ce{CsHSO4} (Figure \ref{fig::method_Ea_msd}c) closely match experimental values \cite{Ishikawa2008, Haile_2007, Blinc_1995, HAYASHI_2004}. We note that experimental uncertainties in activation energies likely exceed the statistical fitting errors derived from our ML MD simulations, primarily due to the limited experimental temperature range (typically 230--260$^\circ$C). Additionally, the deformability of superprotonic \ce{CsH2PO4} \cite{Boysen2009} can lead to changes in sample dimension over time, further increasing the experimental uncertainty.

To assess the error associated with the constant-volume approximation, we performed auxiliary simulations of \ce{CsH2PO4} that incorporated thermal expansion effects \cite{Ikeda_2014}. Over the temperature range of interest, the computed diffusivities are statistically identical (Figure S3, Supporting Information). Consequently, we retain the fixed-volume results as a robust baseline for investigating transport mechanisms. Our simulations accurately reproduce the activation barriers and predict higher proton conductivity in \ce{CsHSO4} than in \ce{CsH2PO4}, consistent with pulsed-field gradient (PFG) NMR diffusion measurements (Figure S4, Supporting Information). Detailed methodology for diffusion coefficient calculations and error analysis is provided in Supporting Information.

\subsection{Two types of rotational motions}

\begin{figure*}[htbp!]
\centering
\includegraphics[width=\linewidth]{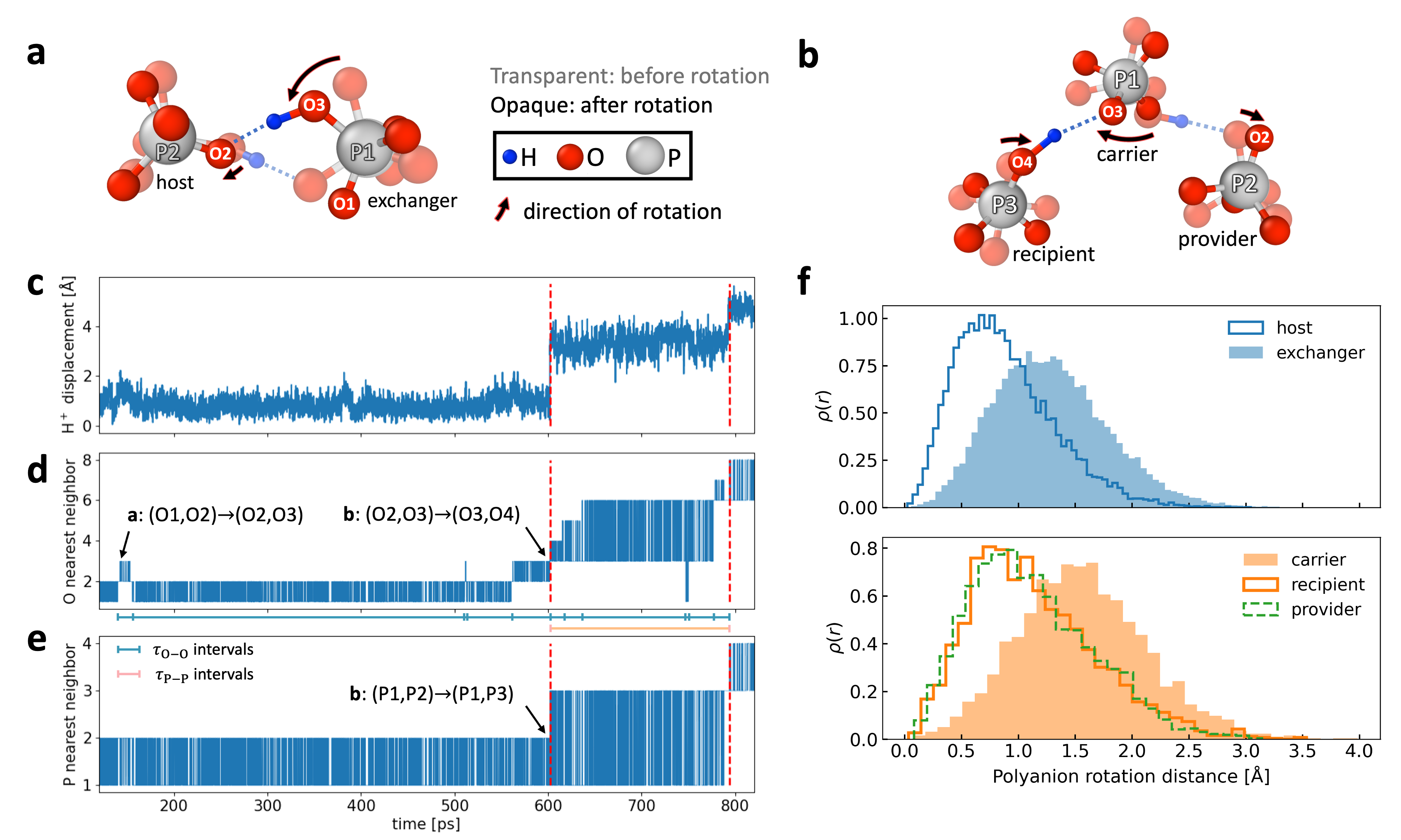}
\caption{\textbf{Single proton trajectory and two types of polyanion rotations.} Illustration of two types of polyanion rotation for a given proton: \textbf{(a)} Nonproductive rotation: the proton resides in the same pair of polyanions after the breaking of \ce{O-H\bond{...}O} bond. \textbf{(b)} Productive rotation: the proton is transferred to a new polyanion assisted by a carrier rotation and experiences a significant long-range motion, labeled by vertical red dashed lines. For both rotation types, the carrier maintains the \ce{O-H} bond during the rotation. An excerpt of a single proton trajectory from a ML MD of \ce{CsH2PO4} at 525 K including: \textbf{(c)} Proton displacement relative to $t=0$. The rapid change of the nearest oxygen neighbor indicates proton rattling within an O-pair. \textbf{(d)} Time evolution of the index of the nearest O neighbor. \textbf{(e)} Time evolution of the index of the P atom defining the \ce{PO4} polyanion containing the nearest O. \textbf{(f)} Comparison of rotational magnitudes between nonproductive (top) and productive (bottom) rotation mechanisms. Shaded histograms represent the rotating polyanions (``exchanger" and ``carrier"). In nonproductive events, the proton-carrying ``host" remains mostly stationary while the ``exchanger" rotates. In productive events, the proton-carrying ``carrier" itself undergoes rotation.}
\label{fig::pair_sw}
\end{figure*}

Building on the finding that overall proton transport properties are accurately predicted by our ML MD simulations and agree with experiments, we proceed to investigate the mechanistic relationship between polyanion rotations and proton transport.

For every proton, we track its nearest O neighbor and observe that the proton rapidly hops between a pair of O anions belonging to different polyanion groups, as shown in the example of \ce{CsH2PO4} (Figure \ref{fig::pair_sw}d). Over the simulation time from 150 ps to 520 ps, the tracked proton jumps back and forth between oxygen atoms labeled as O1 and O2. This process, which we refer to as proton rattling, occurs at characteristic timescales of 0.27 ps for \ce{CsH2PO4} at 525 K and 2.3 ps for \ce{CsHSO4} at 450 K, consistent with previous AIMD studies of these materials \cite{Dressler2020, Lee2008, Wood2007}. However, rattling motions do not directly contribute to long-range transport, as protons remain confined within the same O-pair, where each oxygen is bonded to the phosphorus of a distinct polyanion. Figure \ref{fig::pair_sw}c demonstrates that rattling produces only minor fluctuations in proton displacement. The time interval during which the proton remains associated with a single O-O pair, i.e., the residence time of a given hydrogen bond, is designated $\tau_{\text{O-O}}$.

On timescales that exceed simple rattling events, protons change their nearest neighbor environment through two distinct polyanion rotation mechanisms (Figure \ref{fig::pair_sw}a, b). In what we term \textbf{nonproductive rotations}, a polyanion group (the ``exchanger") rotates, switching the hydrogen bond from one of its oxygen ligands to another. In the example shown in Figure \ref{fig::pair_sw}a, the initial hydrogen bond that was between O2 and O1 is transformed into a new bond between O2 and O3. Because O1 and O3 are ligands of the same exchanger, the proton remains localized between the same pair of polyanion groups (the host and exchanger). Consequently, this change in local environment does not induce significant proton displacement.

In \textbf{productive rotations} (Figure \ref{fig::pair_sw}b), significant proton displacement occurs because the proton remains covalently bonded to an oxygen ligand of the rotating polyanion, termed the ``carrier". As the carrier rotates, it transports the proton from the ``provider" polyanion group to a ``recipient" polyanion group. This process alters the local environment of the proton, changing not only one of its oxygen neighbors but also one of its phosphorus neighbors. For the example shown, at the time indicated by the first set of red dashed lines in Figure \ref{fig::pair_sw}c,d,e, the oxygen nearest neighbors change from (O2, O3) to (O3, O4) and the phosphorus nearest neighbors change from (P1, P2) to (P1, P3). An analogous switching event is indicated by the second set of lines in the figure. This three-polyanion mechanism enables long-range proton diffusion. The time interval during which the proton remains associated with one P-P pair is designated $\tau_{\text{P-P}}$, the residence time of a phosphate-phosphate linkage. The question of whether other oxygen ligands of the exchanger polyanion (Figure \ref{fig::pair_sw}a) induce productive transport is addressed in Section \ref{sec::independence}.

To characterize the magnitude of the underlying polyanion rotations, we quantify the displacement of proton-bonded O anions, $\Delta r_{\text{O}}$, during the rotation period [$t_s, t_f$]. This period is defined from the onset of rotation ($t_s$), when the proton forms a covalent \ce{O-H} bond with either host or carrier polyanion, to its completion ($t_f$), when that bond breaks and the proton resumes rattling. For the exchanger polyanion (Figure \ref{fig::pair_sw}a), we measure the displacement on its oxygen atom O3 that bonds to the proton post-rotation. Both types of rotations are rapid, typically completing within picoseconds (Figure S6, Supporting Information), which corresponds to the fast timescale of the proton hop itself.

Histograms in Figure \ref{fig::pair_sw}f reveal that the largest polyanion motions are associated with carrier rotations, which exhibit greater displacements than the exchanger motions in nonproductive events. Overall, the displacement distributions are similar for the provider and recipient polyanions in productive rotations and the host polyanion in nonproductive ones, though the productive events feature a tail of large displacements absent in nonproductive rotations. These findings highlight the importance of significant carrier rotation for long-range proton hopping.

Although carrier rotation makes the largest contribution to proton displacement, the extent of rotation is modest. The average oxygen displacement of 1.6 Å corresponds to a \ce{P-O} bond rotation of only 62$^{\circ}$ (using $d_{\text{P-O}}=1.55$ Å), indicating that large-scale tumbling of the \ce{PO4} groups is not required for transport. This result extends previous AIMD-based hypotheses \cite{Lee2008} to the nanosecond timescales accessible to our ML MD simulations.

\subsection{Independence of rotational dynamics} \label{sec::independence}

Given the significant magnitude of nonproductive exchanger rotations, it is reasonable to question whether they participate in productive rotations for other protons. To address this, we examine temporal overlaps between rotation periods of these processes in \ce{CsH2PO4} (details in Supporting Information).

The analysis reveals that nonproductive rotations operate largely independently of productive transport. The temporal overlap between them is minimal: 94\% of nonproductive exchanger and 98\% of host rotations occur without a simultaneous productive rotation in the same polyanion (Table S2, Supporting Information). Specifically, when the polyanion undergoes a nonproductive rotation for one proton, it rarely participates simultaneously in a productive rotation for another. These results confirm the statistical independence of the rotational dynamics and justify treating productive and nonproductive rotations as distinct processes.

\subsection{Rates and energy barriers of rotations} \label{sec::CDP_rate_Ea}

\begin{figure*}[htbp!]
    \centering
    \includegraphics[width=\textwidth]{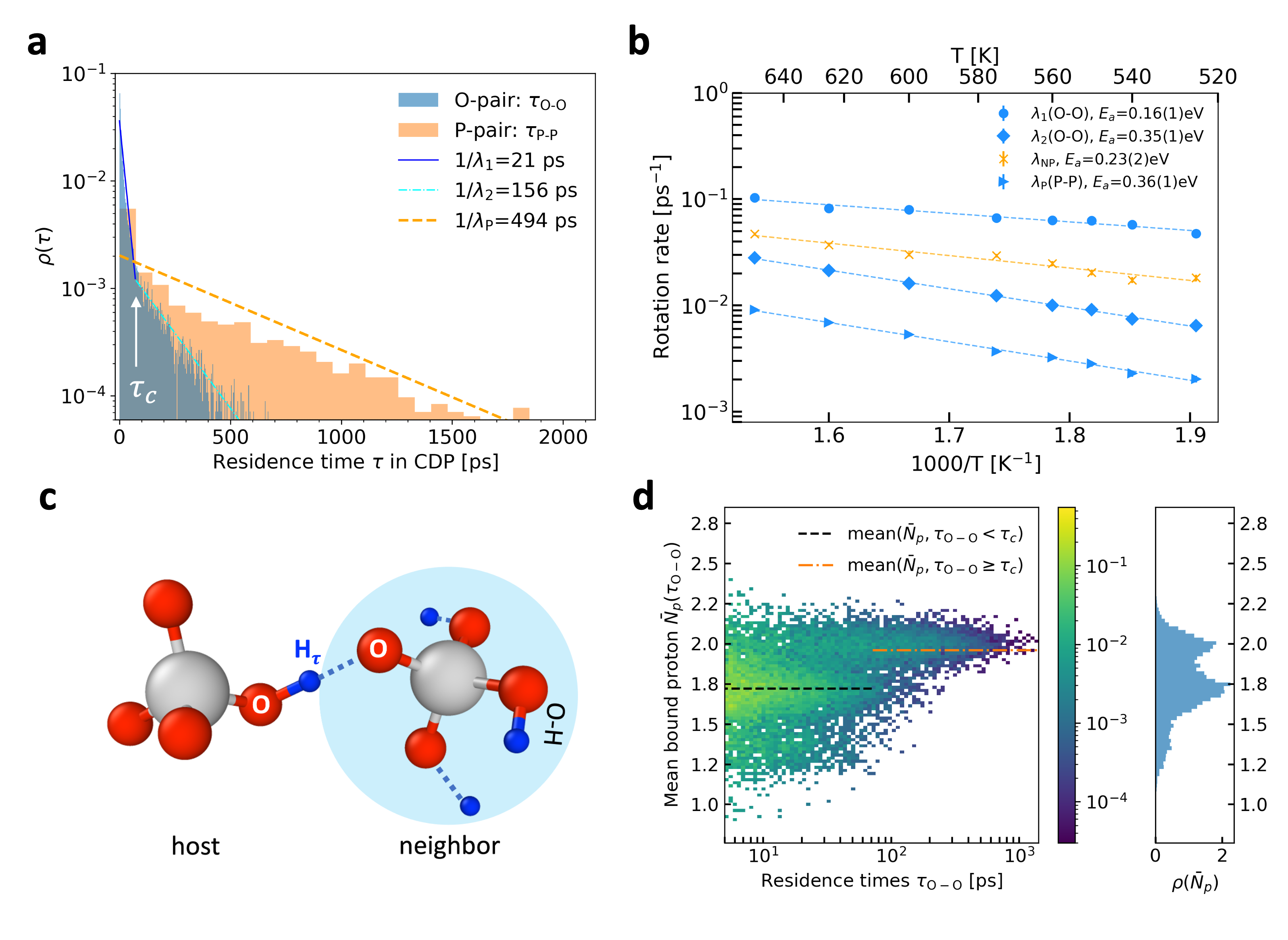}
    \caption{\textbf{Two-rate behavior and effect of bound protons} \textbf{(a)} Distribution of O-pair and P-pair residence times from all protons in a 4 ns ML MD simulation of \ce{CsH2PO4} at 525 K. O-pair residence times include intervals between both productive and nonproductive rotations, exhibiting fast ($\lambda_1$) and slow ($\lambda_2$) rates, while P-pair residence times encompass only intervals between productive rotations, occurring at rate $\lambda_{\text{P}}$. $\tau_c$ marks the transition from fast to slow rotation.
    \textbf{(b)} Arrhenius plot of overall rotation rates ($\lambda_1$, $\lambda_2$), productive rotation rate ($\lambda_{\text{P}}$), and model-derived nonproductive rotation rates ($\lambda_{\text{NP}}$) across various temperatures. Error bars on the symbols indicate the standard error from linear regression fits.
    \textbf{(c)} Counting of bound protons (with \ce{O-H} bond), $N_p(t)$, on the neighbor polyanion (right) at $t \in \tau_{\text{O-O}}$. A neighbor polyanion is defined as the one primarily forming the \ce{H_{\tau}\bond{...}O} bond with its residing proton (H$_\tau$) during $\tau_{\text{O-O}}$. 
    \textbf{(d)} Joint probability density of $\bar{N}_p (\tau_{\text{O-O}})$ and $\tau_{\text{O-O}}$, where $\bar{N}_p(\tau_{\text{O-O}})$ is the time-averaged bound protons number on the neighbor polyanion as a function of $\tau_{\text{O-O}}$. The color scale represents the magnitude of the joint probability. The marginal probability density of $\bar{N}_p$ (right) reveals two distinct peaks in the population of bound protons attached on the neighbor polyanion in \ce{CsH2PO4}. 
    }
    \label{fig::residence_time_PR}
\end{figure*}

To quantify the kinetics of these distinct rotational events, we measure two specific proton residence times, as illustrated in Figure \ref{fig::pair_sw}d,e. The overall rotation rate is estimated by tracking the proton residence time within a given oxygen pair ($\tau_{\text{O-O}}$), defined as the interval between rotation events (terminated by either a productive or nonproductive rotation). In contrast, the productive rotation rate, which governs long-range transport, is isolated by tracking the proton residence time within a phosphate pair ($\tau_{\text{P-P}}$), representing the interval between productive rotations.

Leveraging large-scale ML MD simulations allows us to analyze the full distributions of these residence times (Figure \ref{fig::residence_time_PR}a). The exponential nature of these distributions confirms that rotation events are independent and follow Poisson statistics. The overall rotation interval ($\tau_\text{O-O}$) in \ce{CsH2PO4} exhibits a two-rate decay characterized by a fast rate ($\lambda_1$) and a slow rate ($\lambda_2$), with transition at characteristic time $\tau_c$. Interestingly, detailed decomposition reveals that these fast and slow dynamics are intrinsic to both nonproductive and productive rotations (Figure S7, Supporting Information). Nevertheless, the residence distribution profile of productive rotations ($\tau_\text{P-P}$) is largely captured using a single-rate decay (designated $\lambda_{\text{P}}$). This allows us to isolate the primary timescale governing productive transport.

To isolate nonproductive kinetics, we model the rotation dynamics as a sequence of independent events, where each rotation is either productive (``success") with probability $p$ or nonproductive (``failure") with probability $1-p$. If the events are indeed independent, the number of nonproductive rotations occurring within a single productive interval follows a geometric distribution, and fitting the count data to this distribution yields $p$. The agreement of the measured distribution with a geometric distribution (see Supporting Information Section 14) indicates that each rotation event is statistically independent, meaning that the probability of a productive rotation depends only on the current state rather than the history of prior events. Moreover, the analysis reveals a surprisingly low efficiency for proton transfer: only 10\% of rotations are productive at 525 K, with a weak temperature dependence (which rises to just 16\% at 650 K). Using the relationship $p = \lambda_{\text{P}} / (\lambda_{\text{P}} + \lambda_{\text{NP}})$, we obtain the underlying nonproductive rate via $\lambda_{\text{NP}} = \lambda_{\text{P}}(1-p)/p$.

Having decoupled the rates of the productive and nonproductive rotations, we are now positioned to compare their values and energetics to the empirical observations. In Figure \ref{fig::residence_time_PR}b we show the Arrhenius behavior of four rates, the empirical overall fits ($\lambda_1, \lambda_2$) and the physically isolated productive ($\lambda_{\text{P}}$) and nonproductive ($\lambda_{\text{NP}}$) rates. The productive rotation ($\lambda_{\text{P}}$) exhibits an activation energy of 0.36 eV, which is statistically identical to the empirical slow rate $\lambda_2$ (0.35 eV). This agreement confirms that the slow regime of the overall rotations is indeed dominated by productive events, which in turn, correspond to the rate-limiting step of the long-range diffusion process (0.43 eV, Figure \ref{fig::method_Ea_msd}c).

In contrast, the model-derived nonproductive rate ($\lambda_{\text{NP}}$) yields an activation energy of 0.23 eV, significantly lower than that of productive rotations. These nonproductive events are energetically favorable and occur with much higher frequency than productive rotations. We note that the nonproductive rate $\lambda_{\text{NP}}$ is slower than the empirical fast rate $\lambda_1$ ($E_a=0.16$ eV) by a factor of approximately two. This discrepancy arises because of their differing definitions: $\lambda_1$ captures the fastest intrinsic motions dominating the short-time decay of $\tau_{\text{O-O}}$, whereas $\lambda_{\text{NP}}$ is a discrete event rate derived from counting statistics. Crucially, however, both $\lambda_{\text{NP}}$ and $\lambda_1$ consistently identify a rapid, low-barrier rotational regime that is distinct from the transport-limiting productive events. Kinetic parameters obtained by directly computing the global average rotation frequency (total events per unit time per proton) yielded rates consistent with our model-derived values (see Supporting Information).

The ability to distinguish the activation energy of productive rotations from the combined barrier represents a significant advance. Previous experimental studies were limited to activation energies in the monoclinic phase \cite{Kim2015}, while prior simulations \cite{Wood2007, Dressler2020, Jinnouchi2022} were restricted to pure rotational dynamics via \ce{P-O} bond vector decorrelation times (Eq. \ref{eqn::vec_autocorrelation}). As those methods could not resolve the specific proton-polyanion coupling, our ML MD analysis provides the first quantitative, mechanistic barrier for productive rotations specifically within the superprotonic phase.

Our computed timescales show excellent agreement with experimental data. Using parameters from \textsuperscript{1}H NMR and conductivity measurements \cite{Ishikawa2008}, the correlation time of protonic motion ($\tau_{\text{NMR}}$) is estimated to be 333 ps at 525 K. This falls directly between our computed slow ($\lambda_2^{-1}$) and productive ($\lambda_{\text{P}}^{-1}$) timescales. Furthermore, the productive timescale aligns with the $\sim$500 ps equilibration period required for MD to enter the diffusive regime, highlighting the necessity of ML MD for probing these slow dynamics.

Finally, the discovery of intrinsic two-rate behavior provides a unified view of rotation dynamics in \ce{CsH2PO4}. Two-rate rotation behavior has been proposed previously in the low-temperature monoclinic phase on the basis of NMR studies. By modeling C3 rotations about the \ce{P-O} bonds, Kim et al. \cite{Kim2015} obtained a better agreement with the \textsuperscript{17}O NMR line shapes using two distinct rates rather than a single rate. Our findings reveal that multi-rate character persists in the superprotonic phase. As discussed below, these two rates originate from differences in the local proton-bonding environment, specifically the distinct configurations arising from the two covalently bonded protons.

\subsection{Population of bound protons on rotation rates} \label{sec::bound protons}

To investigate how local proton coordination influences rotation rates, we analyze the bound protons surrounding polyanion pairs during residence time intervals ($\tau_{\text{O-O}}$). Figure \ref{fig::residence_time_PR}c shows a typical configuration where the residing proton H$_\tau$ forms an \ce{O-H} bond with the host polyanion (left) and an \ce{H\bond{...}O} hydrogen bond with the neighbor polyanion (right). We count bound protons ($N_p(t)$) on the neighbor polyanion using a mutual nearest-neighbor criterion, and calculate the time-averaged number $\bar{N}_p (\tau_{\text{O-O}})$: 
\begin{equation}
    \bar{N}_p (\tau_{\text{O-O}}) = \frac{1}{M} \sum_{t=t_i}^{t=t_i+\tau_{\text{O-O}}} N_p(t),
    \label{eqn::mean_proton_coord}
\end{equation}
where $t_i$ is the starting time and $M$ is the number of MD snapshots within $\tau_{\text{O-O}}$. 

The probability density map of $(\bar{N}_p, \tau_{\text{O-O}})$ (Figure \ref{fig::residence_time_PR}d) reveals that shorter residence times ($\tau_{\text{O-O}} < \tau_c$, where $\tau_c$ is defined in Figure \ref{fig::residence_time_PR}) correlate with a reduced number of bound protons ($\bar{N}_p \approx 1.8$), whereas longer residence times correspond to fully bound environments ($\bar{N}_p \approx 2.0$ for $\tau_{\text{O-O}} \geq \tau_c$). This transient decrease in bound protons lowers the rotational barrier, enabling the faster ($\lambda_1$) rotation process. The marginal probability of $\bar{N}_p$ reveals two distinct peaks, indicating substantial local fluctuations in bound proton populations that split rotation rates in \ce{CsH2PO4}. Because the rapidly rotating polyanion group statistically hosts fewer covalently bound protons, these fast orientations are nonproductive for long-range proton transport.

\subsection{\texorpdfstring{\ce{O-H}}{O-H} bond reorientation and slingshot migration} \label{sec::OH}

\begin{figure*}[htbp!]
\centering
\includegraphics[width=\linewidth]{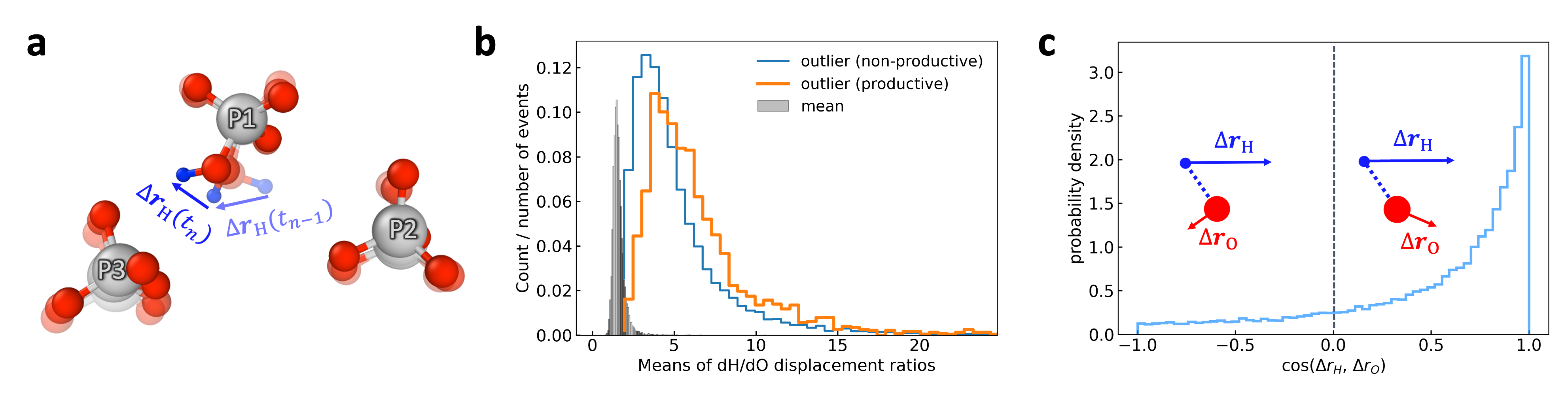}
\caption{\textbf{\ce{O-H} bond reorientation} 
\textbf{(a)} Three consecutive ML MD snapshots illustrating significant \ce{O-H} bond reorientation occurring on the carrier P1, despite minimal rotation of the carrier itself at the moment. 
\textbf{(b)} Distributions of mean dH/dO ratios (defined in Eq. \ref{eqn::HO_ratio}) for both productive and nonproductive rotations in \ce{CsH2PO4}, along with outlier distributions for each.  
\textbf{(c)} Distribution of cosine angles between instantaneous H and O displacements during all rotational motions, illustrating that most \ce{O-H} reorientations extends proton displacement in the direction of the polyanion's rotation.}
\label{fig::OHR}
\end{figure*}

To relate these rotational dynamics to the net transport, we analyzed the distribution of proton displacements that occur during the productive rotation events (Figure S10a, Supporting Information). The analysis shows an average hopping distance of 3.16 Å. This value is physically reasonable, as it is smaller than the 3.5 Å separation between the adjacent proton sites located at the cubic face centers (assuming a lattice parameter of $a = 4.96$ Å). Ishikawa \textit{et al.} estimated the proton jump distance associated with polyhedral rotations in \ce{CsH2PO4} using PFG-NMR experiments \cite{Ishikawa2008} in combination with the activation energy for proton transport ($E_a$) derived from AC impedance measurements (0.39 eV) and obtained a value of 2.7 Å. While the experiments revealed important aspects of the transport behavior, the estimate of the jump distance by this method has a large uncertainty due to its high sensitive to the value of $E_a$. A variation of just $\pm$ 0.02 eV in $E_a$ expands the experimental range in jump distance to 2.11--3.29 Å, a range that includes the computation result of 3.16 Å obtained here.

Taking the true hop distance to be 3.16 Å reveals an important mechanistic discrepancy: polyanion rotation alone cannot account for such a large displacement. A simple geometric calculation, assuming a rigid and linear 2.55 Å \ce{P-O-H} bond (using $d_{\text{O-H}}=1.0$ Å), shows that a 62$^{\circ}$ rotation would displace the proton by only ~2.63 Å. While this nominally coincides with the experimental value of 2.7 Å, as noted above, there is greater uncertainty with the experiment than with the direct computational evaluation. The difference between the rotational displacement (2.63 Å) and the full hopping distance (3.16 Å) implies that proton transfer operates via a cooperative mechanism: the polyanion rotation must be coupled with a secondary motion. This supplementary motion involves \ce{O-H} bond reorientation, constituting the ``proton slingshot" mechanism.

To understand this cooperative motion, we investigate \ce{O-H} bond reorientation during rotation periods. Figure \ref{fig::OHR}a demonstrates how significant \ce{O-H} reorientation on the carrier polyanion enables additional proton displacement toward the next \ce{PO4} polyanion.

We quantify \ce{O-H} reorientation during rotation period $[t_s, t_f]$ using the dH/dO displacement ratio:
\begin{equation}
    \frac{\Delta r_{\text{H}}(t_n)}{\Delta r_{\text{O}}(t_n)} = 
    \frac{|\vb*{r}_{\text{H}}(t_n) - \vb*{r}_{\text{H}}(t_{n-1})|}{|\vb*{r}_{\text{O}}(t_n) - \vb*{r}_{\text{O}}(t_{n-1})|} , \forall t_n, t_{n-1} \in [t_s,t_f]
\label{eqn::HO_ratio}
\end{equation}
where consecutive MD snapshots ($t_n$, $t_{n-1}$) are separated by 50 fs, defined by the trajectory dump frequency. This ratio measures instantaneous proton displacement relative to its bonded O anion, with a value of 1 indicating synchronized movement.

During rotation periods, dH/dO ratios typically fluctuate around 1, with occasional spikes indicating \ce{O-H} reorientation events (Fig. S11, Supporting Information). To characterize these dynamics, we compute two metrics for each rotation period. First, we calculate the baseline \textbf{mean} ratio by averaging all instantaneous dH/dO values within the period. Second, to capture significant reorientation events, we identify spikes where the ratio exceeds a Z-score of 2 and define the \textbf{outlier} value as the average of these high-magnitude ratios. 

Aggregating these metrics across all rotation events yields three distinct distributions (Fig. \ref{fig::OHR}b). The distribution of mean ratios, combined for both rotation types, peaks at 1.5, reflecting baseline thermal fluctuations. In contrast, the distributions of outlier ratios exhibit significantly larger values, peaking at 3.6 for nonproductive rotations and 4.1 for productive rotations. This clear separation between baseline and outlier distributions confirms that substantial \ce{O-H} reorientation accompanies polyanion rotation, with productive events exhibiting slightly larger reorientation magnitudes.

To further evaluate the transport contribution of these reorientations, we analyze the alignment between proton and oxygen displacements through the cosine angle, $\cos \left(\Delta \vb*{r}_{\text{H}}(t_n), \Delta \vb*{r}_{\text{O}}(t_n)\right)$, during rotation periods. Figure \ref{fig::OHR}c shows predominantly aligned motions, leading us to term this the ``proton slingshot" mechanism. This synergistic combination of productive carrier rotation with aligned \ce{O-H} reorientation drives the long-range proton transfer mechanism.

\subsection{Correlation of O-sharing and proton dynamics}

\begin{figure*}[htbp!]
\centering
\includegraphics[width=0.8\linewidth]{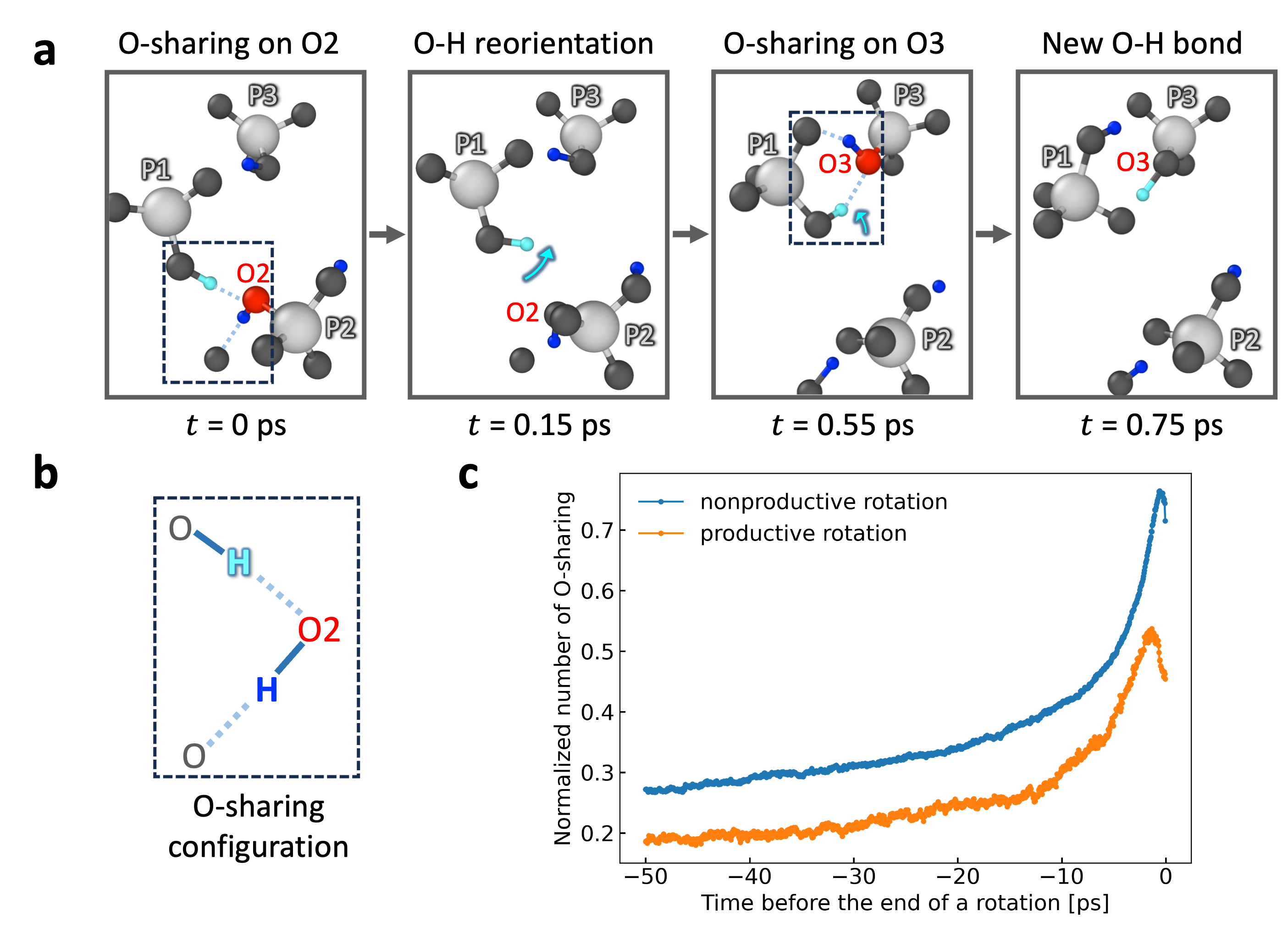}
\caption{\textbf{O-sharing} 
\textbf{(a)} ML MD snapshots illustrating two O-sharing configurations during a productive carrier rotation. The configuration is boxed with shared O anion marked red. At 0 ps, the shared O2 repels the \ce{O-H} bond, reorienting it away from provider P2. At 0.55 ps, carrier P1 brings the hopping proton (cyan) near O3 via rotation, creating another O-sharing configuration. At 0.75 ps, O3 offloads its previously bonded proton (blue) and receives the hopping proton by forming a new \ce{O-H} bond. 
\textbf{(b)} O-sharing configuration, where O2 is simultaneously in the oxygen pairs of two different protons (cyan and blue).
\textbf{(c)} Fraction of rotation events exhibiting active O-sharing as a function of time before rotation completion ($t = 0$). For each rotation event involving a given proton, we track whether the proton is in an O-sharing configuration (panel b) at each time step leading up to completion. All events are time-aligned to their endpoint. Productive rotations (orange, proton hops to a new phosphate pair) and nonproductive rotations (blue, proton remains within the same phosphate pair) are shown separately. 
}
\label{fig::Oshare}
\end{figure*}

Having established the importance of \ce{O-H} reorientation in the proton transfer mechanism, we now examine a unique phenomenon in \ce{CsH2PO4} that further influences proton dynamics during polyanion rotations. This phenomenon arises from the high proton coordination per polyanion in this material.

In \ce{CsH2PO4}, the hydrogen-to-oxygen stoichiometry ensures each proton associates with a distinct O-pair. Our ML MD simulations confirm that two protons occupying the same O-pair is effectively prohibited, with zero instances observed. However, we observe O-sharing events in which a single oxygen atom simultaneously participates in the O-pairs of two different protons (Figure \ref{fig::Oshare}b). At 525 K, an average of 12.8\% of protons participate in such O-sharing configurations at each MD step.

Figure \ref{fig::Oshare}a illustrates O-sharing during a productive rotation. The evolution from $t=0$ to 0.15 ps shows that O-sharing precedes \ce{O-H} bond reorientation and proton hopping, creating local electrostatic repulsion that facilitates the subsequent proton transfer to a new phosphate pair. To quantify the correlation between O-sharing and proton dynamics, we monitor the fraction of rotation events exhibiting active O-sharing from 50 ps before rotation completion (Figure \ref{fig::Oshare}c). The distributions reveal that O-sharing probability increases significantly as both productive and nonproductive rotations approach completion, then decline sharply just prior to the completion of the rotations. The decline corresponds to the formation of a new stable \ce{O-H\bond{...}O} bond as the system approaches its post-transfer equilibrium. These features demonstrate the critical role of O-sharing events in enabling proton transfer in \ce{CsH2PO4}.

Notably, no O-sharing configurations are observed in the \ce{CsHSO4} simulations described below. This distinction reflects the higher proton concentration in \ce{CsH2PO4}, which enables O-sharing events to accompany efficient proton transport.

\subsection{Local proton transfer in \texorpdfstring{\ce{CsHSO4}}{CsHSO4}}

As noted, \ce{CsHSO4} is structurally distinct from \ce{CsH2PO4}, and previous AIMD studies \cite{Lee2008, Dressler2020} have revealed faster polyanion rotations but slower local proton transfer in \ce{CsHSO4} compared to \ce{CsH2PO4}. However, the underlying mechanism behind the faster \ce{SO4} rotations remains unexplored due to the limited spatial scale of AIMD. The present ML MD approach provides an opportunity to uncover these mechanisms.

Taking an analogous, proton-centric approach as applied to \ce{CsH2PO4}, we track the displacements, nearest oxygen neighbors, and nearest polyanion neighbors to a given proton (Figure S13a, Supporting Information). The analysis reveals several distinctions from \ce{CsH2PO4}. First, proton rattling within O-pairs is less frequent in \ce{CsHSO4}, occurring on an average timescale of 1.9 ps at 500 K compared to 0.27 ps at 525 K in \ce{CsH2PO4}. This suggests a deeper double-well potential that inhibits rapid local proton transfer between nearest oxygen neighbors in \ce{CsHSO4}. Second, unlike in \ce{CsH2PO4}, productive rotations in \ce{CsHSO4} usually complete after significant proton displacement has occurred. This sequence delays the hopping step because protons must first reach toward recipient polyanions to form \ce{O-H\bond{...}O} bonds before breaking their \ce{O-H} bonds with carrier polyanion groups and transferring to the recipients (Figure S13b, Supporting Information).

\subsection{Polyanion rotation and \texorpdfstring{\ce{O-H}}{O-H} reorientation in \texorpdfstring{\ce{CsHSO4}}{CsHSO4}}

\begin{figure*}[htbp!]
    \centering
    \includegraphics[width=\textwidth]{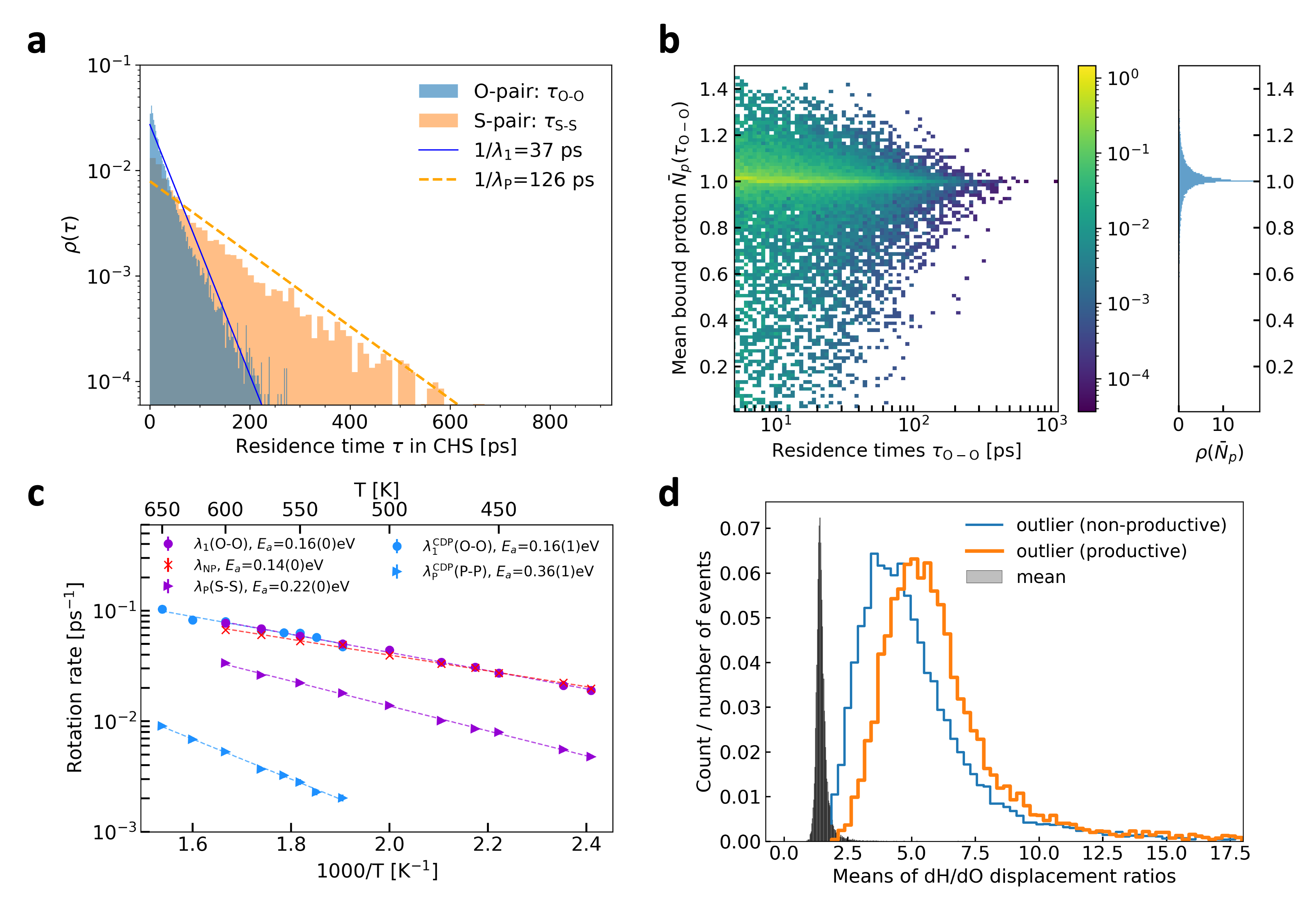}
    \caption{\textbf{Proton slingshot mechanism in \ce{CsHSO4}} 
    \textbf{(a)} Residence time distributions from a 4ns ML MD simulation of \ce{CsHSO4} at 450 K. Both O-pair and S-pair residence time distributions indicate a single-rate polyanion rotation. 
    \textbf{(b)} Joint probability density of $\bar{N}_p (\tau_{\text{O-O}})$ and $\tau_{\text{O-O}}$ at 450 K with marginal probability of $\bar{N}_p$ (right) suggests a single peak in the population of bound protons on the neighbor polyanion of \ce{CsHSO4}. 
    \textbf{(c)} Arrhenius plot of overall ($\lambda_1$), productive ($\lambda_{\text{P}}$), and model-derived nonproductive ($\lambda_{\text{NP}}$) rotation rates in \ce{CsHSO4}. The fast ($\lambda_1$) and productive ($\lambda_{\text{P}}$) rotation rates for \ce{CsH2PO4} are plotted (blue) for comparison. Error bars on the symbols indicate the standard error from linear regression fits.
    \textbf{(d)} Distributions of mean dH/dO displacement ratios from both rotation types, along with outlier distributions for each in \ce{CsHSO4}.}
    \label{fig::CHS_compare}
\end{figure*}

We examine the coupling between polyanion and proton dynamics in \ce{CsHSO4} using the analysis technique from section \ref{sec::CDP_rate_Ea}. We estimate the general polyanion rotation rate from the O-pair residence time ($\tau_{\text{O-O}}$) distribution and the productive rotation rate from the S-pair residence times ($\tau_{\text{S-S}}$). Figure \ref{fig::CHS_compare}a reveals that, unlike the two-rate behavior in \ce{CsH2PO4}, \ce{CsHSO4} exhibits a single characteristic rate for polyanion rotations, as shown in the exponential distribution of $\tau_{\text{O-O}}$. Moreover, productive rotations in \ce{CsHSO4} at 450 K, as given by $\lambda_{\text{P}}$, occur about four times faster than in \ce{CsH2PO4} at 525 K. The timescale of productive rotations in \ce{CsHSO4} aligns with the proton diffusion timescale observed from the MSD profiles (Figure S5, Supporting Information). This indicates that, as with \ce{CsH2PO4}, productive polyanion rotations are closely coupled to long-range proton diffusion.

Proton coordination analysis of \ce{CsHSO4} (Figure \ref{fig::CHS_compare}b) reveals a single peak in the bound proton population ($\bar{N}_p (\tau_{\text{O-O}}) \approx 1$). This uniform environment is consistent with the observation of a single-rate rotation and aligns with previous findings that proton bonding minimally affects polyanion dynamics in this material \cite{Munch_1995}. This behavior contrasts with the bi-modal rate distribution in \ce{CsH2PO4}, which arises from the presence of ``overbonded" and ``underbonded" polyanion groups. The comparison of rates in the Arrhenius plot (Figure \ref{fig::CHS_compare}c) reflects this simpler landscape: \ce{CsHSO4} exhibits a single overall rotation rate ($\lambda_1$) that is identical to the fast rotation rate in \ce{CsH2PO4}, sharing the same activation barrier of 0.16 eV. This suggests that the intrinsic barrier for rotation is similar in both materials when unhindered by surrounding bound protons. However, as already noted, the additional protons in \ce{CsH2PO4} introduce local population fluctuations that create higher activation barriers for slow rotations ($\lambda_2$ in Figure \ref{fig::residence_time_PR}b). As a result of fewer proton-polyanion constraints, \ce{CsHSO4} exhibits more frequent productive rotations (larger $\lambda_{\text{P}}$) than those in \ce{CsH2PO4}, consistent with its higher proton mobility observed in ML MD simulations (Figure \ref{fig::method_Ea_msd}c). Finally, applying the geometric distribution analysis (as in Section \ref{sec::CDP_rate_Ea}) reveals that the nonproductive rate ($\lambda_{\text{NP}}$) in \ce{CsHSO4} closely matches the overall rotation rate ($\lambda_1$). This occurs because the macroscopic rotation rate in \ce{CsHSO4} is, as with \ce{CsH2PO4}, dominated by frequent nonproductive events.

We further investigated whether nonproductive rotations in \ce{CsHSO4} are mechanistically coupled to the productive transport of neighboring protons by analyzing temporal overlaps. We find that the probability of a nonproductive rotation occurring without any simultaneous productive event is $\approx 80\%$. While this ``no-overlap" fraction is slightly lower than that observed in \ce{CsH2PO4} ($>90\%$), we attribute this difference to timescale disparities rather than increased coupling. Because the covalent \ce{O-H} bond in \ce{CsHSO4} breaks less frequently, the computed average rotation period is ten times longer than in \ce{CsH2PO4}. Since our overlap metric is based on temporal intersection, these expanded duration windows naturally increase the statistical probability of an overlap occurring, even if the events remain mechanistically independent. Thus, we conclude that nonproductive rotations in \ce{CsHSO4} do not significantly induce productive transport in neighbors.

Analysis of the proton hopping distance in \ce{CsHSO4} reveals a discrepancy similar to that found in \ce{CsH2PO4}: the proton displacement exceeds the distance covered by \ce{SO4} rotation alone. The average proton displacement during productive rotations in our simulations is 2.49 Å at 450 K (Figure S10b, Supporting Information). This computed displacement is consistent with the $\approx$ 2.3 Å proton jump measured by QENS \cite{Belushkin_1992}. The oxygen atoms undergo a mean displacement of only 1.14 Å during the rotation of a carrier polyanion group (Figure S14, Supporting Information). Assuming a rigid \ce{HSO4^-} geometry ($d_{\text{S-O}}=1.44$ Å, $d_{\text{O-H}}=1.0$ Å), the theoretical proton displacement associated with this rotation would be limited to just 1.93 Å. The difference between this rigid rotation value and the observed 2.49 Å hop implies a necessary contribution from \ce{O-H} reorientation, which we quantify below.

As with \ce{CsH2PO4}, significant \ce{O-H} reorientation occurs in \ce{CsHSO4} during sulfate rotations (Figure \ref{fig::CHS_compare}d). The mean dH/dO ratios peak around 1.4, while outlier distributions reach 3.7 in nonproductive rotations and 5.2 in productive rotations. Again, similar to \ce{CsH2PO4}, the direction of \ce{O-H} reorientation aligns with the carrier rotation (Figure S12, Supporting Information), highlighting its role in facilitating proton jumps. Our finding that long-range proton transport is always accompanied by \ce{O-H} reorientation in two compounds with very different structures suggests that the ``slingshot" mechanism is a common feature of superprotonic transport. These findings substantiate a hypothesis loosely hinted at by Baranov \textit{et al.} \cite{Baranov_2003} in superprotonic \ce{CsHSO4}.

\subsection{Polyanion orientations in \texorpdfstring{\ce{CsHSO4}}{CsHSO4} and \texorpdfstring{\ce{CsH2PO4}}{CsH2PO4}}

\begin{figure*}[htbp!]
\centering
\includegraphics[width=\linewidth]{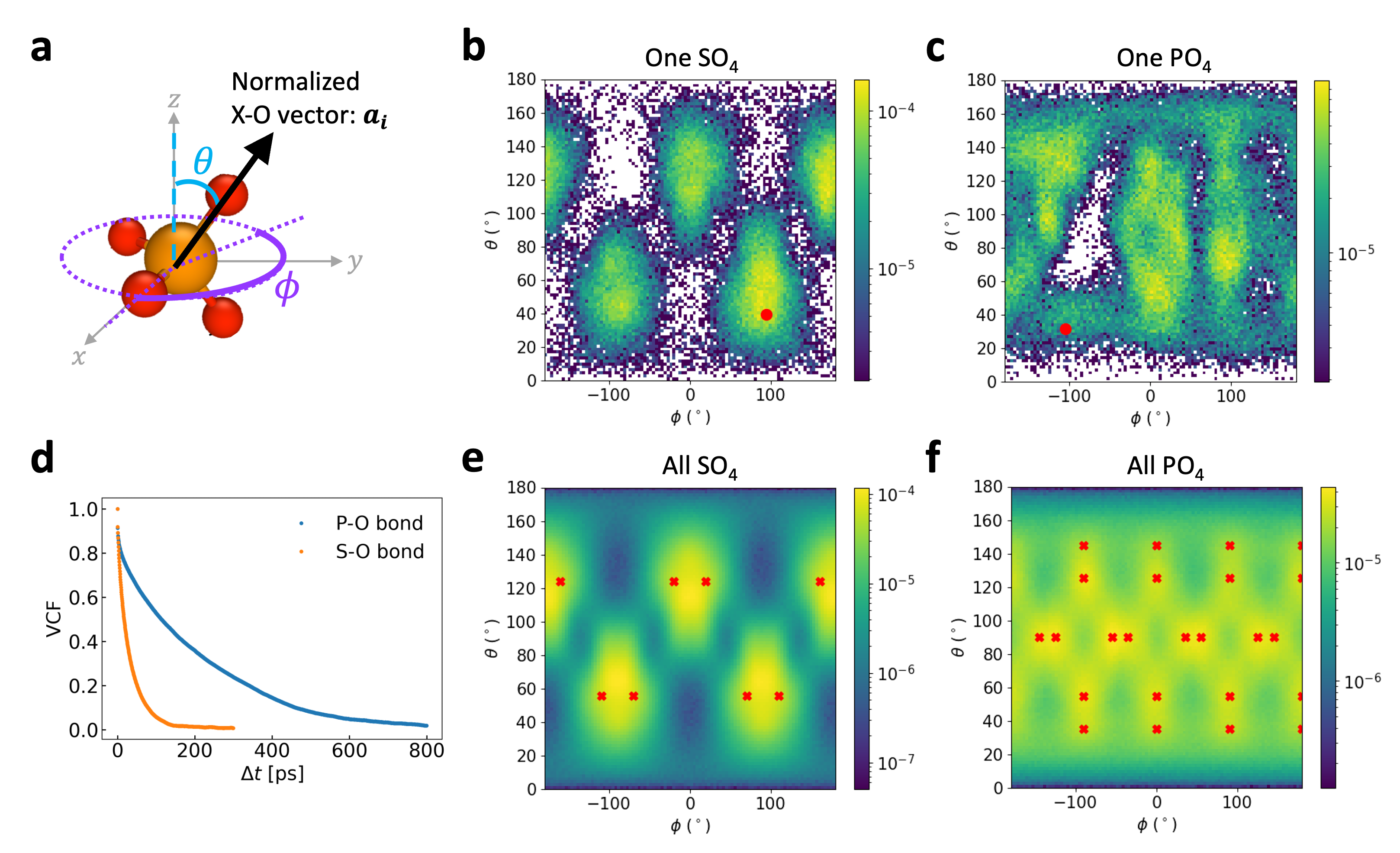}
\caption{\textbf{Polyanion orientations in \ce{CsHSO4} and \ce{CsH2PO4}} \textbf{a} Illustration of a normalized \ce{X-O} vector in an \ce{XO4} polyanion and its spherical angles $(\phi, \theta)$. Cumulative probability distribution of the spherical angles of a single \ce{X-O} vector in one selected polyanion for \textbf{b} \ce{CsHSO4} and \textbf{c} \ce{CsH2PO4} from their respective 4 ns ML MDs at 550 K. Each red dot indicates the initial spherical angles of the tracked \ce{X-O} bond. \textbf{d} VCF calculated by Eq. \ref{eqn::vec_autocorrelation} for \ce{CsHSO4} and \ce{CsH2PO4} from 4 ns ML MDs at 550K. Cumulative distribution of the spherical angles of the \ce{X-O} bond with the minimum initial $\theta$ from all polyanions for \textbf{e} \ce{CsHSO4} and \textbf{f} \ce{CsH2PO4}. Red crosses denote experimental oxygen sites \cite{Jirak_1987, Yamada_2004}. \ce{SO4} polyanions exhibit more distinct and well-defined orientations compared to \ce{PO4} polyanions. }
\label{fig::XO4_config}
\end{figure*}

We further evaluate and compare \ce{SO4} and \ce{PO4} dynamics by analyzing their preferred orientations and the timescales required to explore the rotational configuration space. In this analysis, the orientation of a polyanion group is defined by the alignment of a constituent X-O bond vector, parameterized by azimuthal ($\phi$) and polar ($\theta$) angles relative to the lattice vectors (Figure \ref{fig::XO4_config}a). Specifically, $\theta=0^{\circ}$ corresponds to an X-O bond aligned with the $c$-axis, while $\phi=0^{\circ}, \theta=90^{\circ}$ corresponds to alignment with the $a$-axis. Tracking this vector for an arbitrarily selected polyanion group for a period of 4 ns yields cumulative orientation distributions shown in Figure \ref{fig::XO4_config}b,c. \ce{SO4} exhibits distinct peaks, indicating preferences for specific orientations, whereas \ce{PO4} displays rotational frustration and a more disordered distribution. 

This orientational difference is further reflected by the probability distributions across the full orientation space, Figure \ref{fig::XO4_config}e,f (details in Supporting Information). The \ce{SO4} orientations in \ce{CsHSO4} display clear, well-defined peaks that align with experimental oxygen sites (marked by red crosses), whereas \ce{PO4} orientations in \ce{CsH2PO4} show broader, flatter distribution profiles, though the peaks again match experimental orientations.

Counterintuitively, the \ce{PO4} group, linked by four hydrogen bonds, exhibits greater freedom to probe a wider range of orientations than the \ce{SO4} group, which is linked by only two. This result coincides with the higher site symmetry of the \ce{PO4} group in \ce{CsH2PO4} as compared to the site symmetry of the \ce{SO4} group in \ce{CsHSO4}. 

The rate at which polyanions decouple from a given orientation is another valuable metric for evaluating the dynamics of solid acid systems. We characterize this rate according to the time-window averaged vector autocorrelation function (VCF) \cite{Dressler2020,Jinnouchi2022}:
\begin{equation}
    \mathrm{VCF} (\Delta t) = \frac{1}{M} \frac{1}{N} \sum_{t=0}^{T - \Delta t} \sum_{i=1}^{N} \vb*{a}_i (t) \cdot \vb*{a}_i (t + \Delta t).
    \label{eqn::vec_autocorrelation}
\end{equation}
where $\vb*{a}_i (t)$ is the unit \ce{X-O} vector for the $i$-th polyanion at time $t$, $M$ is the number of time windows with duration $\Delta t$, $N$ is the number of \ce{XO4} polyanions, and $T$ is the total time considered in the calculation.

Figure \ref{fig::XO4_config}d shows that the VCF of \ce{S-O} bond decays to 0.5 at 17 ps, whereas the \ce{P-O} bond requires 114 ps to reach this VCF value at the same simulation temperature, indicating significantly faster \ce{SO4} rotations. Thus, although the sulfate groups in \ce{CsHSO4} experience greater constraints in the orientations they can access, the phosphate groups in \ce{CsH2PO4} remain correlated in their orientation over longer periods of time. It remains to be seen whether this stems from the difference in symmetry (tetragonal vs cubic) or the differing proton concentrations. Moreover, it may be impossible to disentangle structure from chemistry if the latter determines the former. 

\section{Conclusion}

We present a unified paradigm for proton transport in solid acid compounds, specifically \ce{CsH2PO4} and \ce{CsHSO4}. We reveal a ``proton slingshot" mechanism, in which cooperative motion between small-scale polyanion rotation and significant \ce{O-H} bond reorientation facilitates long-range proton transfer. This challenges the conventional Grotthuss model and provides a more nuanced understanding of the transport process.

We further distinguish between productive polyanion rotations, which drive long-range transport, and nonproductive rotations, which do not. In nonproductive rotations, the polyanion group undergoes a rotation, but no protons are covalently attached to any of its oxygen ligands during the rotation. The high frequency of proton rattling across O-O bonds frequently leaves polyanions momentarily free of covalently bonded protons, providing ample opportunity for such rotations. In both \ce{CsH2PO4} and \ce{CsHSO4}, the activation barriers for productive rotations closely align with those of proton diffusion, underscoring the critical role of these rotations in high proton mobility. 

While the transport processes in \ce{CsH2PO4} and \ce{CsHSO4} are similar, our work reveals distinctions that correlate with the experimentally observed differences in transport behavior. We quantify the overall rotation rate by the frequency at which protons change their nearest oxygen neighbor, and the productive rotation rate by the frequency at which they change their nearest phosphate (or sulfate) neighbor. In \ce{CsH2PO4}, we observe two time-scales for phosphate rotation: a slower rate corresponding to productive rotations, and a faster rate that occurs when the number of covalently bonded protons falls below the average of two. In contrast, \ce{CsHSO4} exhibits a single time-scale for general sulfate rotation. Although the productive rotation rate in \ce{CsHSO4} is lower than its general rotation rate, it is substantially higher than the productive rate in \ce{CsH2PO4}, by almost an order of magnitude at 525 K. While previous studies have recognized higher polyanion rotation rates in \ce{CsHSO4}, the distinct types of rotations identified here were not previously resolved.

Our work elucidates the impact of proton concentration on rotational dynamics, explaining the observed differences between \ce{CsH2PO4} and \ce{CsHSO4}. The higher proton content in \ce{CsH2PO4} leads to two-rate rotations and increased rotational frustration, contrasting with the single-rate, faster rotations in \ce{CsHSO4}. This insight suggests that reducing the proton concentration could potentially enhance conductivity by alleviating rotational constraints.

The extended spatial and temporal scales of our ML-accelerated MD simulations, combined with our novel analysis techniques, provide access to the full picture of proton transport in these materials. This approach offers a unified framework for characterizing proton and polyanion dynamics in solid acid compounds, opening new avenues for optimizing superprotonic conductors for next-generation fuel cells and energy storage devices.

\section{Data availability}

The \textit{ab initio} dataset and ML MD trajectories of \ce{CsH2PO4} and \ce{CsHSO4} generated in this study are openly available on Zenodo at \url{https://doi.org/10.5281/zenodo.14649022}.

\section{Code availability}
The analysis tools and implementation details will be deposited in the same Zenodo record as the dataset, \url{https://doi.org/10.5281/zenodo.14649022}, upon publication of this article. Open-source software implementations of FLARE and Allegro are available at \url{https://github.com/mir-group/flare} and \url{https://github.com/mir-group/allegro} (Allegro git commit e5dcd1953168fee056970fa7dd86294c3d1f437f).

\section{Author Information}
\subsection{Corresponding Authors}

\textbf{Menghang Wang} -- Harvard John A. Paulson School of Engineering and Applied Sciences, Harvard University, Cambridge, MA 02138, USA; Email: menghangwang@g.harvard.edu

\textbf{Boris Kozinsky} -- Harvard John A. Paulson School of Engineering and Applied Sciences, Harvard University, Cambridge, MA 02138, USA; Email: bkoz@g.harvard.edu

\textbf{Sossina Haile} -- Materials Science and Engineering, Northwestern University, Evanston, IL 60208, USA; Email: sossina.haile@northwestern.edu

\subsection{Authors}

\textbf{Jingxuan Ding} -- Harvard John A. Paulson School of Engineering and Applied Sciences, Harvard University, Cambridge, MA 02138, USA

\textbf{Grace Xiong} -- Materials Science and Engineering, Northwestern University, Evanston, IL 60208, USA

\textbf{Ni Zhan} -- Department of Computer Science, Princeton University, Princeton, NJ 08544, USA

\textbf{Cameron J. Owen} -- Harvard John A. Paulson School of Engineering and Applied Sciences, Harvard University, Cambridge, MA 02138, USA

\textbf{Albert Musaelian} -- Harvard John A. Paulson School of Engineering and Applied Sciences, Harvard University, Cambridge, MA 02138, USA

\textbf{Yu Xie} -- Microsoft Research AI for Science, Berlin, 10178, Germany 

\textbf{Nicola Molinari} -- Robert Bosch Research and Technology Center, Watertown, MA 02472, USA

\textbf{Ryan P. Adams} -- Department of Computer Science, Princeton University, Princeton, NJ 08544, USA

\subsection{Author Contributions}
The manuscript was written through contributions of all authors. All authors have given approval to the final version of the manuscript. 

\section{Competing interests}

The authors declare no competing interest.

\section{Acknowledgements}
The authors gratefully acknowledge Kyle Bystrom for insightful discussions on DFT calculations, Anders Johansson for suggesting the residence time approach and guidance on using LAMMPS, and Gordon Peiker for providing feedback on the manuscript figures. The authors also thank Jenny Hoffman and Suzanne Smith for their editing feedback on the initial version of the manuscript.

This work was primarily supported by National Science Foundation, Office of Advanced Cyberinfrastructure (OAC), under Award No. 2118201. M.W., G.X., N.Z., R.P.A, S.H., and B.K. were supported by National Science Foundation, OAC, under Award No. 2118201. J.D. was supported by the National Science Foundation under Grant No. DMR-2119351. C.J.O was supported by the U.S. Department of Energy, Office of Science, Office of Basic Energy Sciences, Chemical Sciences, Geosciences, and Biosciences Division under Award Number DE-SC0022199. A.M. and B.K. were supported by Harvard University Materials Research Science and Engineering Center under Grant No. DMR-2011754. Y.X was supported by the National Science Foundation under Grant No. OAC-2003725. N.M. was supported by Bosch Research. The authors acknowledge computing resources provided by the Harvard University FAS Division of Science Research Computing Group. The analysis and visualization of MD simulations were conducted using OVITO software \cite{Stukowski2009}. 

\section{References}
\bibliographystyle{naturemag}
\bibliography{pconduct.bib}

\end{document}


\maketitle

\tableofcontents

\clearpage
\section{H position in \texorpdfstring{\ce{CsH2PO4}}{CsH2PO4}} 
\label{SI::H_pos}

For \ce{CsH2PO4}, there are no experimental reports of hydrogen positions in the literature. Therefore, we use the geometry proposed in \cite{Lee2008} to position H atoms in the cubic \ce{CsH2PO4} as the starting structure. Based on Fig. 6 in \cite{Lee2008}, the following assumptions were made: the P, O, and H atoms are on the same $z$-plane, the \ce{P-O-H} angle is 110°, and the \ce{O-H} intra-phosphate group distance is 1.02 Å. The other atomic coordinates (Cs, P, O) were sourced from experimentally derived data in \cite{Botez_2007}. For an O atom at (0.2299, 0.6270, 0.5), the calculated position of the H atom is (0.3002, 0.8202, 0.5). 

\section{DFT calculation} \label{SI:DFT_setup}

We utilized plane wave basis sets and projector augmented wave (PAW) in our calculations, with a plane wave cut-off energy set at 700 eV for both \ce{CsH2PO4} and \ce{CsHSO4}. The convergence criterion for the electron self-consistent loop, EDIFF, was set to $10^{-5}$. Given that both \ce{CsH2PO4} and \ce{CsHSO4} are insulators, Gaussian smearing with a width of 0.1 eV was applied. In addition, LASPH was enabled to include non-spherical contributions in the PAW spheres. 

The Monkhorst-Pack scheme was applied for Brillouin zone sampling. For the superprotonic phase (Pm$\bar{3}$m) of \ce{CsH2PO4}, the cubic unit cell \cite{Yamada_2004} contains one formula unit with a lattice parameter of 4.9615 Å. We utilized a $2\times2\times2$ $k$-point mesh for both the $2\times2\times2$ supercell (64 atoms) and the $3\times3\times3$ supercell (144 atoms). For the monoclinic \ce{CsH2PO4} (P2$_1$/m), the unit cell \cite{preisinger_1994} contains two formula units with lattice parameters: a = 7.91 Å, b = 6.38 Å, c = 4.88 Å. We used a $2\times2\times2$ $k$-point mesh for the $2\times2\times2$ supercell (128 atoms). Superprotonic \ce{CsHSO4} (I$4_1$/amd) has a unit cell \cite{Jirak_1987} with four formula units and lattice parameters: a, b = 5.72 Å, c = 14.23 Å. Both monoclinic phases (III and II, both with symmetry of P2$_1$/c) of \ce{CsHSO4} have four formula units each. Phase III \cite{Chisholm2000} has a = 8.21 Å, b = 5.81 Å, c = 10.98 Å, while phase II \cite{Itoh1990} has a = 7.80 Å, b = 8.21 Å, c = 8.04 Å. We used a $2\times2\times2$ $k$-point mesh for the $2\times2\times1$ supercell (112 atoms) for all three phases.

\section{Dataset generation}
Starting structures of superprotonic phases were generated using OrderDisorderedStructureTransformation from \verb|pymatgen| \cite{Ong_2013}, selecting configurations with proper tetrahedral \ce{PO4} and \ce{SO4} geometries. Internal atomic positions were optimized using VASP (ISIF = 2) while maintaining cell parameters, with Cs and P positions additionally constrained for superprotonic \ce{CsH2PO4}. Monoclinic phase structures of both compounds were fully relaxed using VASP (ISIF = 3).

Table \ref{tab::training_data} summarizes the active learning trajectory setup. Each trajectory began with an empty SGP model to maximize diverse atomic environment sampling while avoiding initial data bias and preventing SGP slowdown as the dataset expands. Data collection ended when the frequency of DFT calls reduced to approximately 1 per 1000 MD steps, ensuring comprehensive sampling of high-uncertainty environments. Simulations were conducted above 800 K to enhance configuration diversity through increased kinetic contributions. We employed different stress tensor constraints in LAMMPS \textit{NpT} simulations (triclinic ``tri" and anisotropic ``aniso") to capture varied cell-symmetry effects. The ACE B2 descriptor \cite{Drautz_2019} was configured with a body order of five and varied radial cutoffs. Radial distribution function analysis confirmed that sampled structures remained physically realistic, free from decomposition, melting, or unrealistic bond lengths.

For \ce{CsH2PO4}, we collected 1,196 MD snapshots spanning 300 to 1,000 K, incorporating both superprotonic (Pm$\bar{3}$m) and monoclinic (P2$_1$/m) phases with system sizes of 64, 128, and 144 atoms. The \ce{CsHSO4} dataset comprises 1,636 MD snapshots from 200 to 2,000 K, covering superprotonic (I$4_1$/amd) and monoclinic (III and II) phases. 

\begin{table*}[htbp!]
\centering
\begin{tabular}{ccccccc}
\toprule
Phase & Temp. & Press. & Ensemble & $N_{\text{atoms}}$ & $r_c$ & $N_{\text{DFT}}$ \\ \midrule
\multirow{3}{*}{\shortstack{\ce{CsH2PO4}\\(P2$_1$/m)}} 
    & 300  & 1 & NPT (aniso) & 128 & 5.5& 21 \\ 
    & 400  & 1 & NPT (aniso) & 128 & 5.5& 23 \\ 
    & 500  & 1 & NPT (aniso) & 128 & 5.5& 64\\ \midrule
\multirow{9}{*}{\shortstack{\ce{CsH2PO4}\\(Pm$\bar{3}$m)}} 
    & 600  & 1 & NPT (aniso) & 144 &5.5& 91   \\ 
    & 800  & 1 & NPT (aniso) & 144 &5.5& 102  \\ 
    & 1000 & 1 & NPT (aniso) & 144 &5.5& 94   \\ 
    & 525  & 1 & NPT (tri)   & 64  &5.5& 139  \\  
    & 600  & 1 & NPT (aniso) & 64  &5.5& 115  \\  
    & 700  & 1 & NPT (tri)   & 64  &5.5& 55   \\ 
    & 800  & 1 & NPT (aniso) & 64  &5.5& 167  \\  
    & 1000 & 1 & NPT (tri)   & 64  &5.5& 90   \\
    & 1000 & 1 & NPT (tri)   & 64  &5.5& 235  \\  \midrule
\multirow{6}{*}{\shortstack{\ce{CsHSO4}\\(P2$_1$/c, III)}} 
    & 200         & 0     & NPT (tri) & 112    &5.5& 36            \\ 
    & 300         & 0     & NPT (tri) & 112    &5.5& 35            \\ 
    & 300         & 1     & NPT (tri) & 112    &5.5& 39            \\ 
    & 300         & 1.2e4 & NPT (tri) & 112    &4.0& 12            \\ 
    & 300 to 600  & 0     & NPT (tri) & 112    &5.5& 66            \\ 
    & 400         & 0     & NPT (tri) & 112    &5.5& 37            \\ 
    & 500         & 0     & NPT (tri) & 112    &5.5& 54            \\ 
    & 800         & 0     & NPT (tri) & 112    &5.5& 85            \\ \hline 
\multirow{6}{*}{\shortstack{\ce{CsHSO4}\\(P2$_1$/c, II)}}
    & 360 & 1     & NPT (tri)   & 112 &4.0& 21 \\
    & 360 & 1.2e4 & NPT (tri)   & 112 &4.0& 17 \\
    & 400 & 1     & NPT (aniso) & 112 &5.5& 26 \\
    & 500 & 1     & NPT (tri)   & 112 &5.5& 47 \\
    & 800 & 1     & NPT (tri)   & 112 &5.5& 95 \\
    & 800 & 1     & NPT (aniso) & 112 &5.5& 36 \\ \midrule
\multirow{3}{*}{\shortstack{\ce{CsHSO4}\\(I$4_1$/amd, I)}}
    & 450        & 0   & NPT (tri)   & 112 &5.5& 181 \\
    & 450 to 300 & 0   & NPT (iso)   & 112 &5.5& 84  \\
    & 800        & 0   & NPT (tri)   & 112 &5.5& 104 \\ \midrule
\multirow{8}{*}{\shortstack{\ce{CsHSO4}\\(I$4_1$/amd, I)}}
    & 450        & 0   & NPT (tri)       & 112  &5.5& 48  \\
    & 550        & 0   & NPT (tri)       & 112  &5.5& 37  \\
    & 550        & 0   & NPT (aniso)     & 112  &5.5& 32  \\ 
    & 550        & 1   & NPT (aniso+tri) & 112  &5.5& 55  \\ 
    & 800        & 0   & NPT (tri)       & 112  &5.5& 54  \\
    & 2000       & 0   & NPT (aniso+tri) & 112  &5.5& 80  \\
    & 800        & 1   & NPT (tri)       & 112  &6.6& 165 \\
    & 2000       & 1   & NPT (tri)       & 112  &6.0& 190 \\ \bottomrule
\end{tabular}
\caption{Setup details and dataset information from FLARE Bayesian active learning workflow \cite{Vandermause_2020, Vandermause_2022, Xie_2023}: the material phase; simulation temperature (K);  pressure (Bar);  statistical ensemble;  number of atoms $N_{\text{atoms}}$ in the simulation cell;  radial cutoff $r_c$ (Å) of the ACE descriptor; number of collected MD snapshots $N_{\text{DFT}}$ with DFT-calculated labels.}
\label{tab::training_data}
\end{table*}

\section{MLFF validation}

\begin{figure*}[htbp!]
\centering
\includegraphics[width=\linewidth]{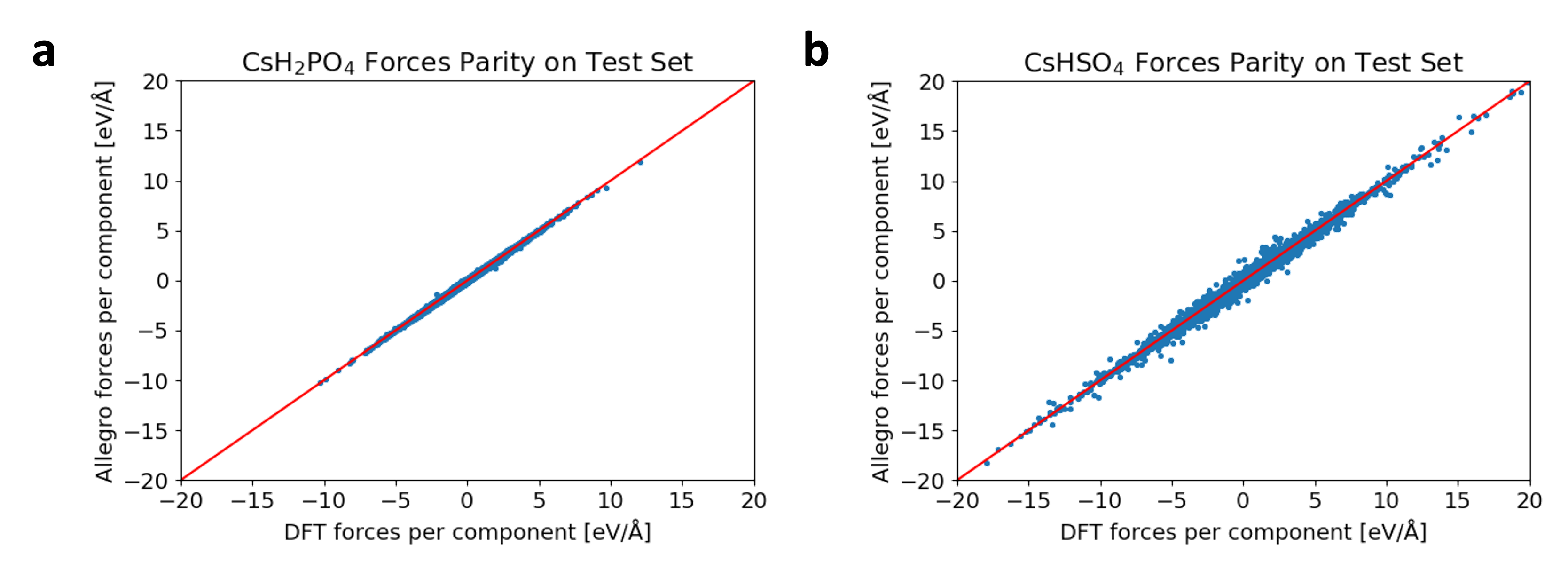}
\caption{Force parity plot on the test set of \textbf{(a)} \ce{CsH2PO4} and \textbf{(b)} \ce{CsHSO4}, respectively.}
\label{fig::force_parity}
\end{figure*}

The root mean square error (RMSE) on forces from test sets is 45 meV/Å for \ce{CsH2PO4} with 130 MD snapshots and 114 meV/Å for \ce{CsHSO4} with 199 MD snapshots. The comparisons between the DFT forces and ML forces of all components are shown in Figure \ref{fig::force_parity}.

\begin{figure*}[htbp!]
    \centering
    \includegraphics[width=\linewidth]{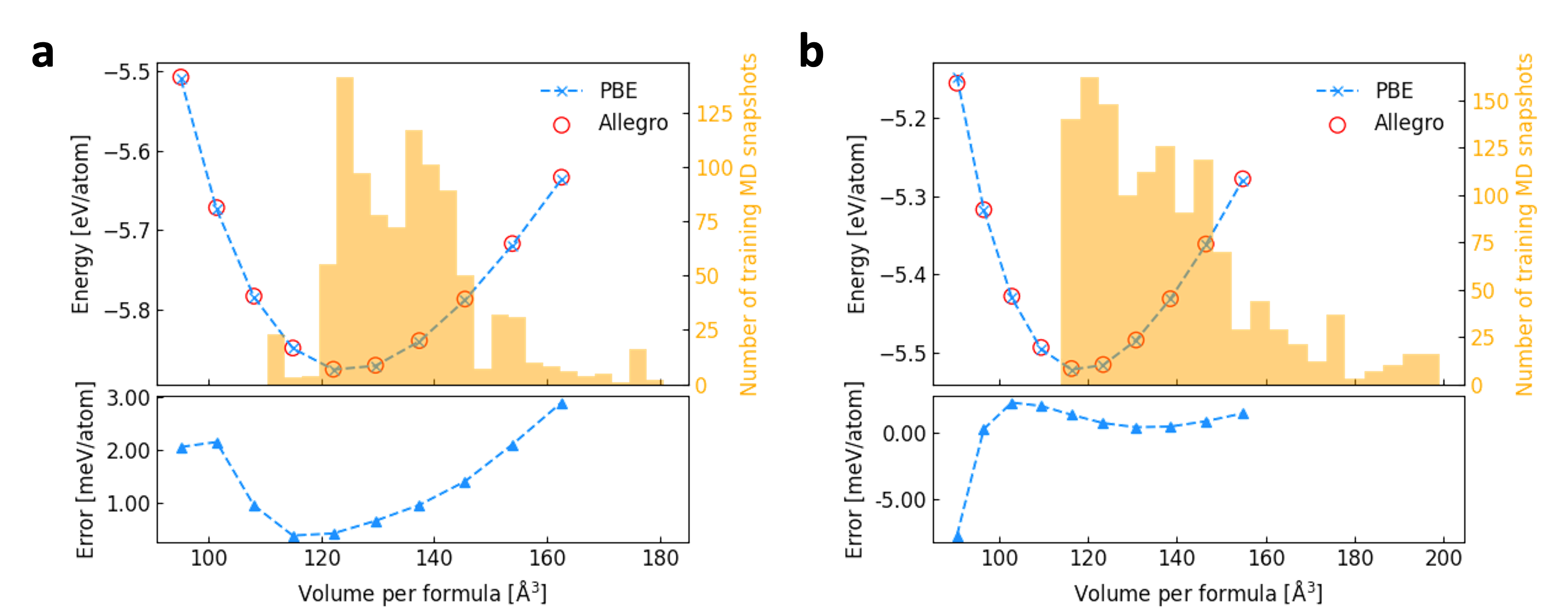}
    \caption{Energy-volume curves for the superprotonic phases of \textbf{(a)} \ce{CsH2PO4} (144 atoms, 525 K) and \textbf{(b)} \ce{CsHSO4} (112 atoms, 500 K) from a snapshot of corresponding NVT AIMD simulations. Data are predicted by the Allegro model (circles) and DFT calculations with PBE functional (crosses). Predictive errors remain within 2 meV/atom for lattice parameter 0.92 to 1.08 for \ce{CsH2PO4} and 0.94 to 1.10 for \ce{CsHSO4}. The orange histogram illustrates the training data distribution relative to volume per formula unit.}
    \label{fig::eqn_of_state}
\end{figure*}

In our equation-of-state test on \ce{CsH2PO4} and \ce{CsHSO4}, we obtained the structures of their superprotonic phases from the final frame of 100 ps NVT AIMD simulations (144 atoms at 525 K for \ce{CsH2PO4}, and 112 atoms at 500 K for \ce{CsHSO4}), which are not included in the training data. We isotropically scaled cell parameters to compare energies predicted by Allegro with those calculated via DFT (PBE). As shown in Figure \ref{fig::eqn_of_state}, Allegro's energy predictions align with DFT within 2 meV/atom for lattice parameter scaling from 0.92 to 1.08 in \ce{CsH2PO4} and 0.94 to 1.10 for \ce{CsHSO4}. The predictive error is defined as $(E_\text{Allegro} - E_\text{PBE})/N_\text{atoms}$, with $N_\text{atoms}$ being the number of atoms in the cell. 

\newpage
\section{ML MD setup}

For \ce{CsH2PO4}, we ran simulations for 4 ns at temperatures of 525, 540, 550, 560, and 575 K, and for 3 ns at temperatures of 600, 625, and 650 K. For \ce{CsHSO4}, we ran simulations for 4 ns at temperatures of 415, 425, 450, 460, 475, 500, 525, 550, 575, and 600 K. We used the same experimental lattice parameters across different temperatures. 

\newpage
\section{Thermal expansion effect on cubic \texorpdfstring{\ce{CsH2PO4}}{CsH2PO4}}

\begin{figure*}[htbp!]
\centering
\includegraphics[width=0.8\linewidth]{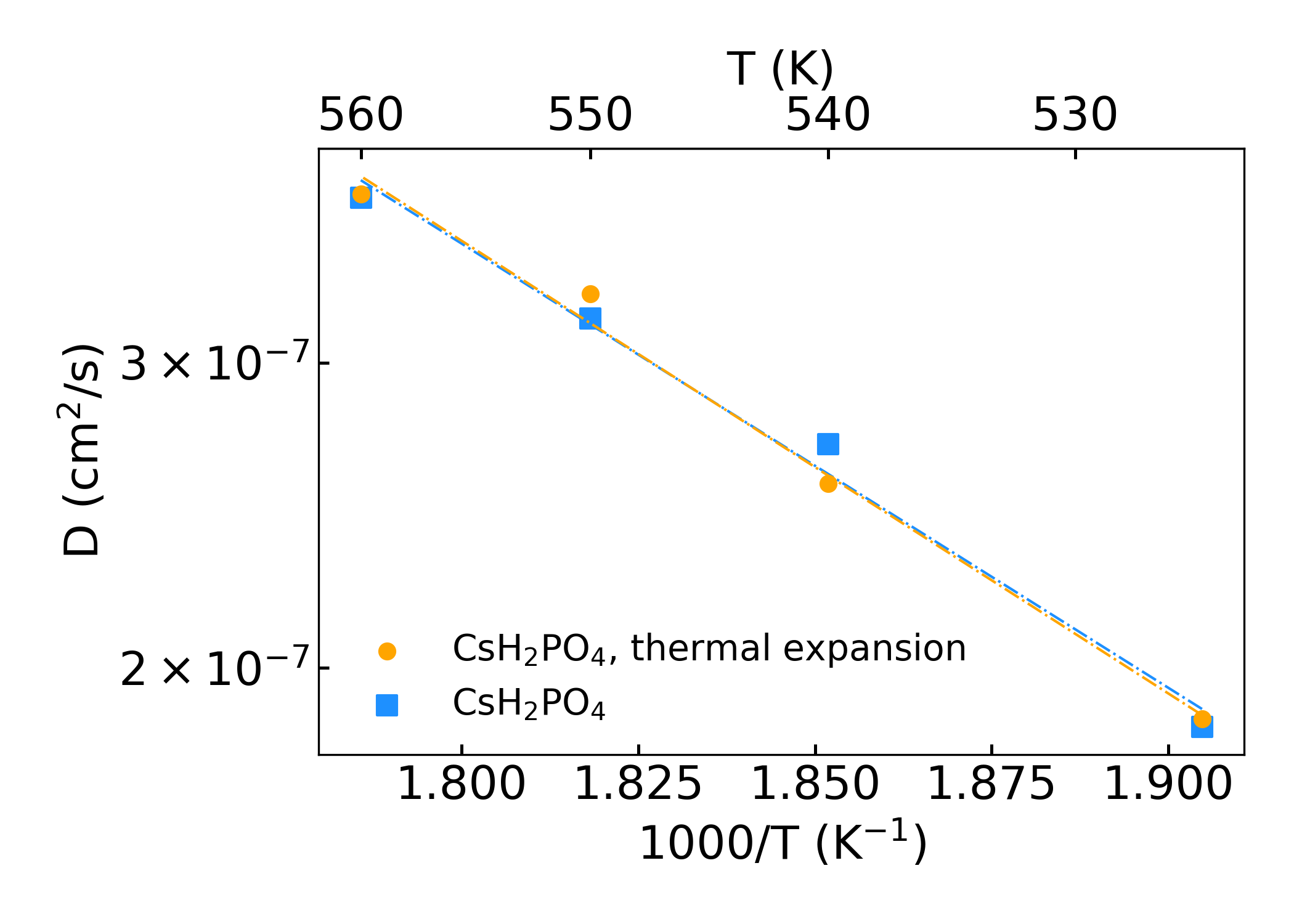}
\caption{Diffusion coefficients computed from simulations using the thermal expansion rates reported in \cite{Ikeda_2014}. Orange markers indicate results from ML MD simulations that include lattice expansion under NVT conditions, while blue markers correspond to simulations using the fixed experimental lattice constant of 4.9615 Å. The mean relative difference between the two datasets is 2.55\%. The simulated temperature range (230--290 $^\circ$C), corresponds to the experimental stability window.}
\label{fig::thermal_expansion}
\end{figure*}

\newpage
\section{Experimental diffusion measurement}

\begin{figure*}[htbp!]
\centering
\includegraphics[width=0.8\linewidth]{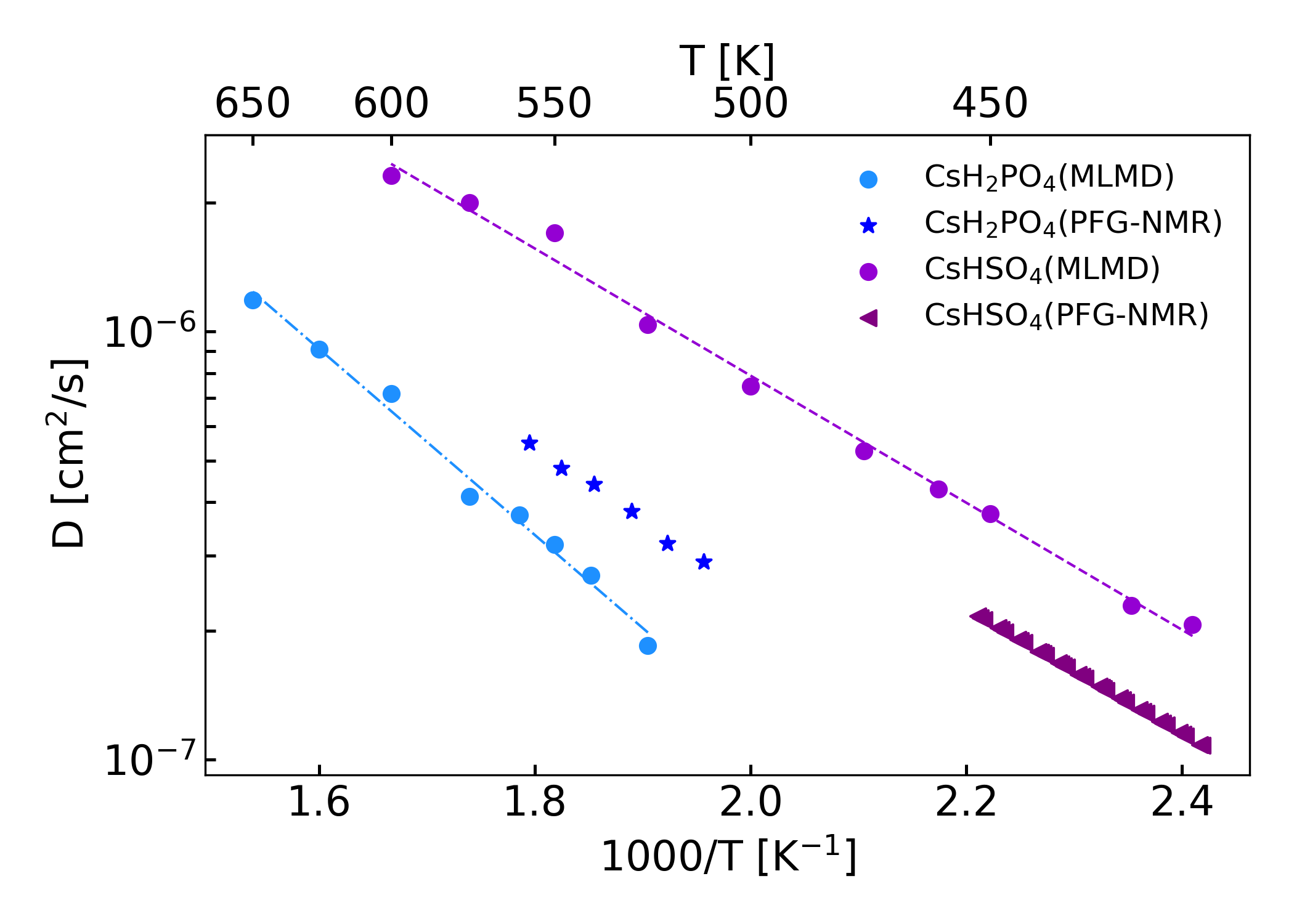}
\caption{Diffusion coefficients and activation energies obtained from pulsed-field gradient NMR (PFG-NMR) and ML MD simulations. PFG-NMR data are taken from Blinc \textit{et al.} \cite{Blinc_1995} for \ce{CsHSO4} and Ishikawa \textit{et al.}\cite{Ishikawa2008} for \ce{CsH2PO4}.}
\label{fig::sim_vs_experiment}
\end{figure*}

\newpage
\section{Diffusion coefficients and error estimation} \label{sec:msd_diffusion}

To obtain the proton diffusion coefficient ($D$), we fitted the slope of the proton MSD in the diffusive regime according to the Einstein equation. The MSD was computed with time window averaging \cite{He2018}, and the curve was truncated at 70\% of the total simulation time to mitigate poor statistics from a few long time-windows. We estimated the slope using a fitting window with $\tau \geq 1$ ns. 

For \ce{CsH2PO4}, we estimated the standard deviation of the fitted diffusion coefficients at selected temperatures using $N$ number of independent ML MD simulations. Each simulation used a different velocity random seed but began with the same initial structure. In the table below, Temp. refers to the temperature of the ML MD simulation, $N$ represents the number of independent trajectories, and std. dev. is the standard deviation of diffusion coefficients across these trajectories. Notably, the standard deviation is an order of magnitude smaller than the calculated diffusion coefficients.
\begin{center}
    \begin{tabular}{ccc} 
     \toprule
     Temp. (K) & $N$ & std. dev. (cm$^2$/s) \\ \midrule
     525 & 5 & 1.013$\times$10$^{-8}$ \\ 
     550 & 10 & 2.023$\times$10$^{-8}$ \\  
     575 & 5 & 2.642$\times$10$^{-8}$ \\ 
     \bottomrule
    \end{tabular}
\end{center}

\section{MSD of \texorpdfstring{\ce{CsHSO4}}{CsHSO4}}

\begin{figure}[htbp!]
\centering
\includegraphics[width=0.7\linewidth]{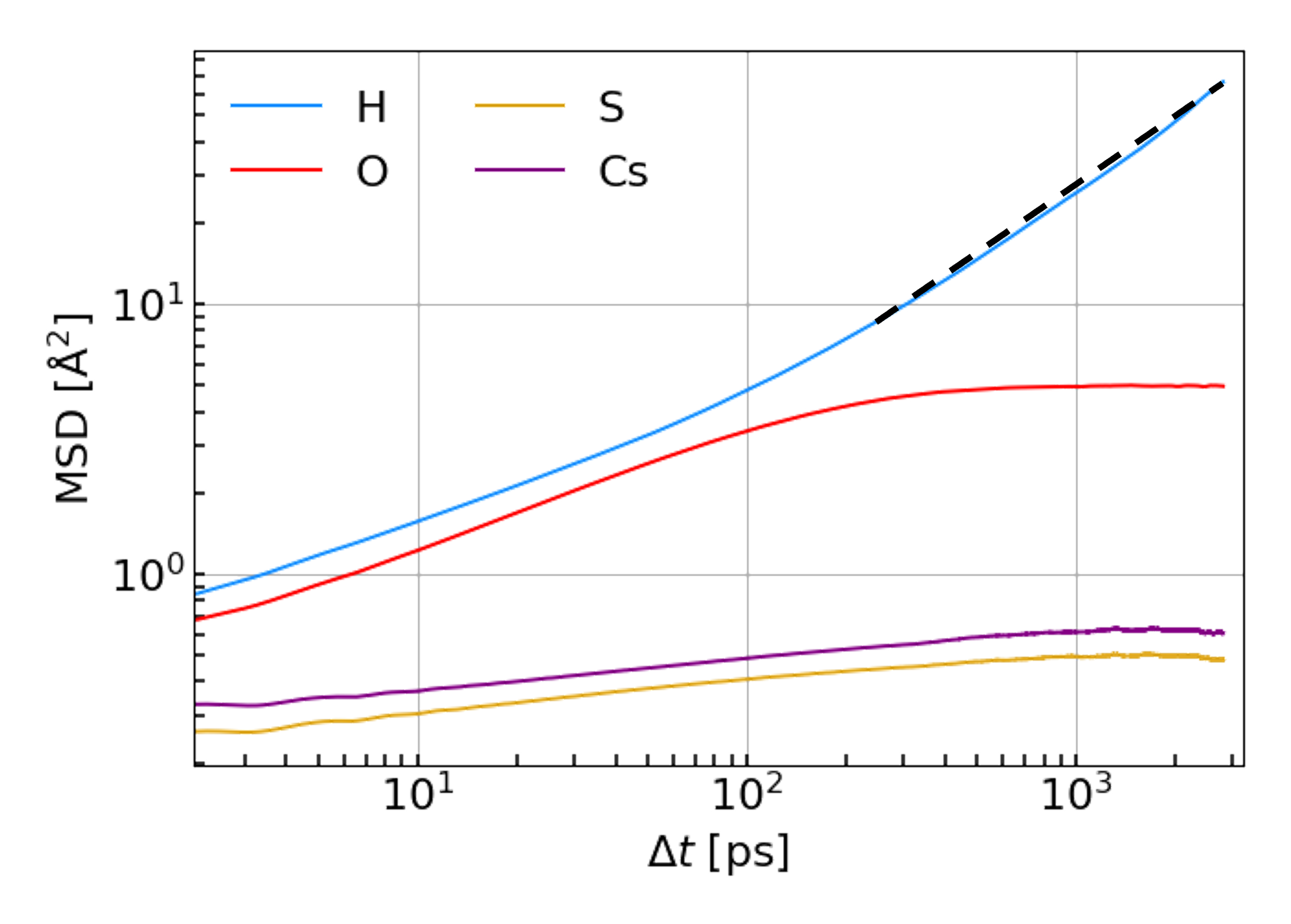}
\caption{MSDs of each species in \ce{CsHSO4} from a 4 ns ML MD at 450 K. Proton diffusive regime starts at 200 ps, and the oxygen MSD plateaus around 500 ps.}
\label{fig::CHS_msd}
\end{figure}

Fig. \ref{fig::CHS_msd} illustrates the significance of long-timescale MD simulations in examining proton and polyanion dynamics in \ce{CsHSO4}.

\newpage
\section{Rotation period}

\begin{figure}[htbp!]
\centering
\includegraphics[width=\linewidth]{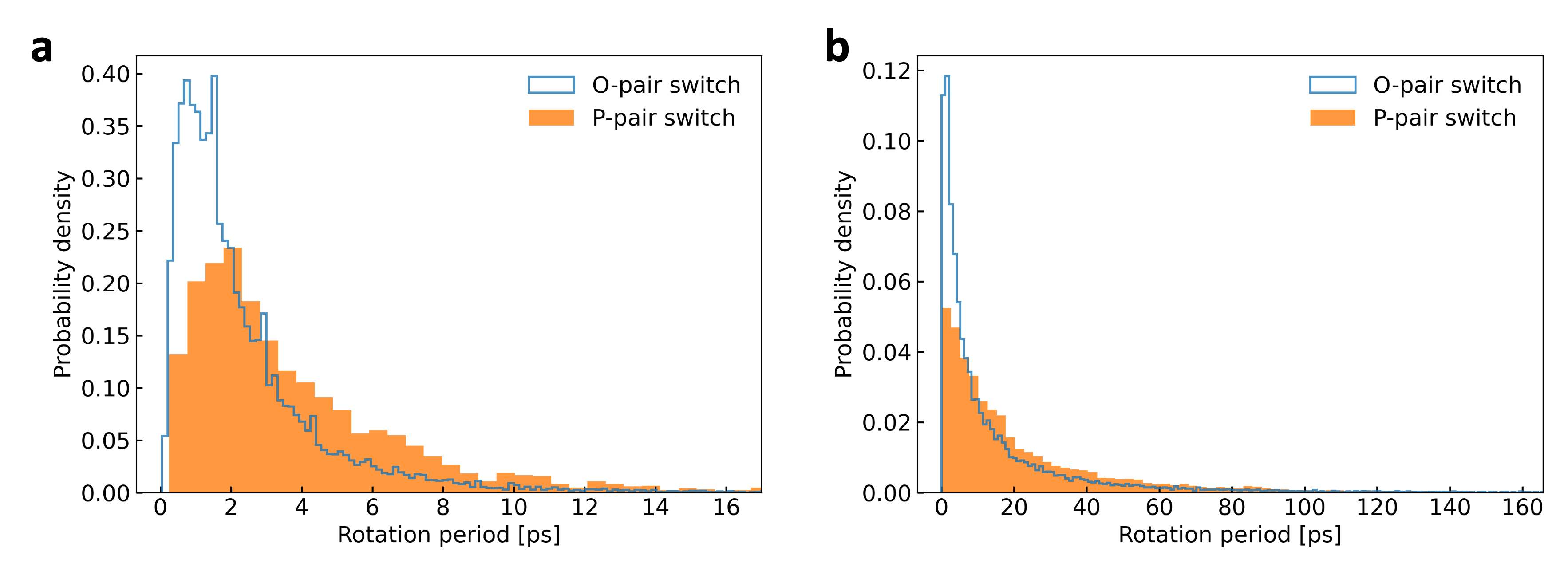}
\caption{Distribution of polyanion rotation periods from ML MD simulations of (a) \ce{CsH2PO4} at 525 K and (b) \ce{CsHSO4} at 450 K.}
\label{fig::rot_period_hist}
\end{figure}

Figure \ref{fig::rot_period_hist} shows the rotation period length ($t_f-t_i$) distributions for \ce{CsH2PO4} and \ce{CsHSO4} from ML MD simulations. In \ce{CsH2PO4}, average rotation periods for O-pair and P-pair switches are 2.4 ps and 3.9 ps, respectively. \ce{CsHSO4} exhibits longer periods (15.8 ps and 24.5 ps) due to slower \ce{O-H} bond breaking. In both compounds, these timescales are significantly shorter than intervals between pair switch events. The x-axes are truncated at the 99th percentile for clarity. Despite differing timescales, these observations support treating polyanion rotation as a discrete event in both materials.

\newpage
\section{Overlaps between nonproductive and productive rotations}

For each rotation, the rotation interval defined by $t_s$ to $t_f$ is assigned to the participating polyanions. Specifically, intervals where a polyanion acts as a host or exchanger are classified as nonproductive, forming the set $\{ (t_{s,\text{NP}}, t_{f,\text{NP}}) \}$. Conversely, intervals where it acts as a provider, carrier, or recipient are classified as productive, forming the set $\{ (t_{s,\text{P}}, t_{f,\text{P}}) \}$.

We define an overlap as any instance where the time intervals intersect:
\begin{equation}
     t_{s,\text{NP}} \leq t_{s,\text{P}} \leq t_{f,\text{NP}} 
     \text{ or } 
     t_{s,\text{NP}} \leq t_{f,\text{P}} \leq t_{f,\text{NP}} .
\end{equation}
This definition provides a strict upper bound on overlap estimates, as it considers the entire duration of the rotation rather than just the instantaneous endpoints of proton transfer events. 

Table \ref{tab::overlap_rot_CDP} summarizes the overlap statistics for \ce{CsH2PO4} at 525 K. The results indicate that overlaps are rare: when a polyanion serves as an exchanger or host in a nonproductive rotation, it is largely decoupled from simultaneous productive roles (overlap $<6\%$ for all cases). This confirms that nonproductive and productive events in \ce{CsH2PO4} can be treated as effectively independent. 

\begin{table*}[htbp!]
\centering
\begin{tabular}{
>{\centering\arraybackslash}p{0.28\linewidth}
>{\centering\arraybackslash}p{0.22\linewidth}
ccc}
\toprule
Role in a nonproductive rotation & Role in a productive rotation & No overlap & 1 overlap & 2 overlaps \\ \midrule
exchanger & carrier   & 99.26\% & 0.73\% & 0.01\%    \\ 
exchanger & recipient & 93.97\% & 5.93\% & 0.1\% \\ 
exchanger & provider  & 93.93\% & 5.99\% & 0.08\% \\ 
host   & carrier   & 98.58\% & 1.38\% & 0.04\%    \\ 
host   & recipient & 97.87\% & 2.11\% & 0.03\%    \\ 
host   & provider  & 97.70\% & 2.28\% & 0.03\% \\  \bottomrule
\end{tabular}
\caption{Overlaps between nonproductive and productive rotations in \ce{CsH2PO4}.}
\label{tab::overlap_rot_CDP}
\end{table*}

In contrast, Table \ref{tab::overlap_rot_CHS} shows a notably higher overlap probability for \ce{CsHSO4} at 450 K. However, this increase does not necessarily imply a stronger mechanistic coupling between rotation types. Instead, it is largely attributable to the differing timescales of rotation periods in the two materials. In \ce{CsHSO4}, the \ce{O-H} bond remains bonded longer during polyanion rotation, resulting in average rotation periods that are an order of magnitude longer than those in \ce{CsH2PO4} (Figure \ref{fig::rot_period_hist}). Since our overlap metric is based on temporal intersection, these longer duration windows naturally increase the statistical probability of an overlap occurring, even if the exchanger rotation is not actively driving the concurrent productive transport event. Even with this biased metric, the overlap remains limited: an exchanger polyanion in a non-productive rotation coincides with a productive role less than 20\% of the time, while a host polyanion overlaps less than 28\% of the time.

\begin{table*}[htbp!]
\centering
\begin{tabular}{
>{\centering\arraybackslash}p{0.24\linewidth}
>{\centering\arraybackslash}p{0.20\linewidth}
cccc}
\toprule
Role in a nonproductive rotation & Role in a productive rotation & No overlap & 1 overlap & 2 overlaps & 3 overlaps \\ \midrule
exchanger & carrier   & 82.29\% & 17.04\% & 0.67\% & 0\% \\ 
exchanger & recipient & 82.28\% & 16.33\% & 1.35\% & 0.04\% \\ 
exchanger & provider  & 81.01\% & 17.52\% & 1.44\% & 0.03\% \\ 
host   & carrier   & 98.01\% & 1.88\%  & 0.11\% & 0\% \\ 
host   & recipient & 73.35\% & 23.43\% & 3.06\% & 0.17\% \\ 
host   & provider  & 72.29\% & 24.36\% & 3.15\% & 0.18\% \\  \bottomrule
\end{tabular}
\caption{Overlaps between nonproductive and productive rotations in \ce{CsHSO4}.}
\label{tab::overlap_rot_CHS}
\end{table*}

\newpage
\section{Residence time analysis} 


We collected the residence time from each proton trajectory and excluded the intervals bounded by either the start or the end of the simulation, which artificially terminate a proton residency. To ensure a robust rate fitting and accurately represent the probability density function, we determined the optimal histogram bin size using the Freedman-Diaconis rule. Furthermore, to exclude the poorly sampled, low-probability tail from the fit, we considered regions of probability density $\rho(\tau_{\mathrm{O-O}}) > 2\times10^{-4}$ for O-pair and $\rho(\tau_{\mathrm{P-P}})> 5\times10^{-5}$ for P-pair residence times.

To identify the transition point $\tau_c$, we minimized the difference in $R^2$ values between segments before and after $\tau_c$, as shown in Eq. \ref{eqn::R-square}, while scanning the values of $\tau_c$ incrementally.
\begin{equation}
    R^2(\tau < \tau_c) - R^2(\tau \geq \tau_c)
    \label{eqn::R-square}
\end{equation}

We then used \verb|curve_fit()| from \verb|scipy.optimize| \cite{scipy} to fit a piecewise linear function with four parameters ($k_1, k_2, \tau_c, b_1$) to the residence time distribution $(\tau, \log(\rho))$:
\begin{equation}
    \begin{cases} 
      k_1 \tau + b_1 & \tau < \tau_c \\
      k_2 (\tau - \tau_c)  + k_1 \tau_c + b_1 & \tau \geq \tau_c
    \end{cases}
\label{eqn::piece-wise}
\end{equation}
where $k_1$ is the fast rate, $k_2$ is the slow rate, and $b_1$ is the vertical intercept with $\log(\rho)$. This approach accurately estimated the transition point $\tau_c$, distinguishing between fast and slow rotations. 

\newpage
\section{Categorized residence time distribution}

\begin{figure}[htbp!]
\centering
\includegraphics[width=0.8\linewidth]{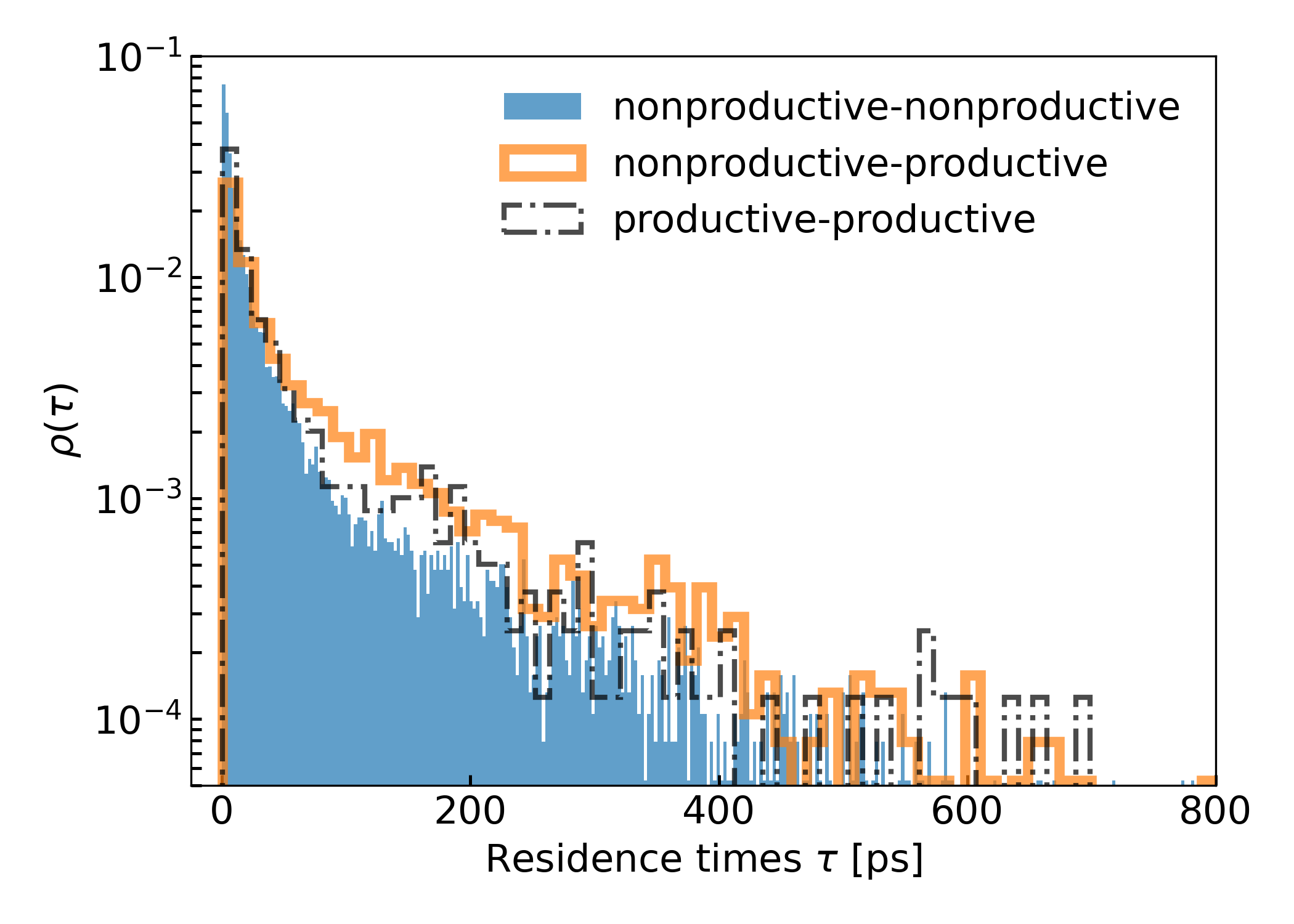}
\caption{O-pair residence times distribution at 525 K in \ce{CsH2PO4} categorized into the intervals separated by two nonproductive rotations (nonproductive-nonproductive), one nonproductive rotation and one productive rotation (nonproductive-productive), and two productive rotations (productive-productive).}
\label{fig::O-pair_types}
\end{figure}

The overall residence times ($\tau_{\text{O-O}}$) can be categorized based on the specific rotation events defining the start and end of the interval, such as intervals bounded by a nonproductive and a productive rotation (nonproductive-productive). Notably, residence times bounded strictly by two consecutive productive rotations, with no intervening nonproductive events, represent a specific subset of the full $\tau_{\text{P-P}}$ distribution. Analyzing this subset in isolation would introduce a systematic statistical bias by selectively excluding longer intervals that contain intermediate nonproductive rotations. Consequently, rates derived from this subset would be artificially skewed toward shorter timescales. As shown in Figure \ref{fig::O-pair_types}, while the two-rate behavior is inherent across both rotation types, including productive ones, we based our primary analysis on the complete $\tau_{\text{P-P}}$ distribution to extract the robust average rate of productive rotation.

\newpage
\section{Estimation of productive rotation probability} \label{si-sec:productive_prob}

To estimate the probability of a productive rotation, we modeled the rotational dynamics as a sequence of independent Bernoulli trials. In this framework, a productive rotation constituted a ``success'' (with probability $p$) and a nonproductive rotation constituted a ``failure'' (with probability $1-p$). Assuming each rotation is an independent event, the number of failures $N$ that occurred before the first success followed a geometric distribution. We emphasized that the distribution of failures $N$ is discrete. The probability mass function for $N$ was:
$$P(N) = (1-p)^N p$$
This distribution decayed exponentially with $N$. Taking the natural logarithm of this equation yielded:
$$\ln(P(N)) = N \ln(1-p) + \ln(p)$$

This equation had the form of a straight line, $y = mx + c$, where $y = \ln(P(N))$, $x = N$, and the slope was $m = \ln(1-p)$. Thus, $p$ could be determined from the slope of a linear regression to the log-probability plot via $p = 1 - e^m$.

\begin{figure}[htbp!]
\centering
\includegraphics[width=\linewidth]{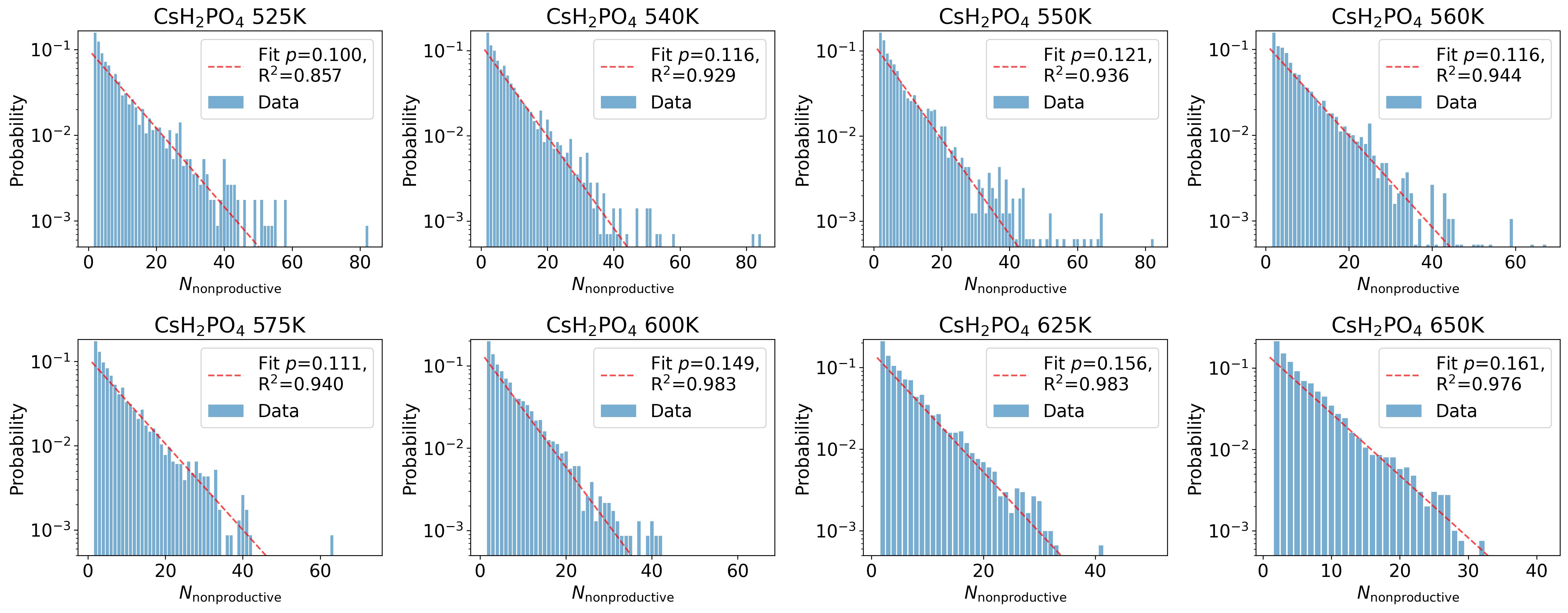}
\includegraphics[width=\linewidth]{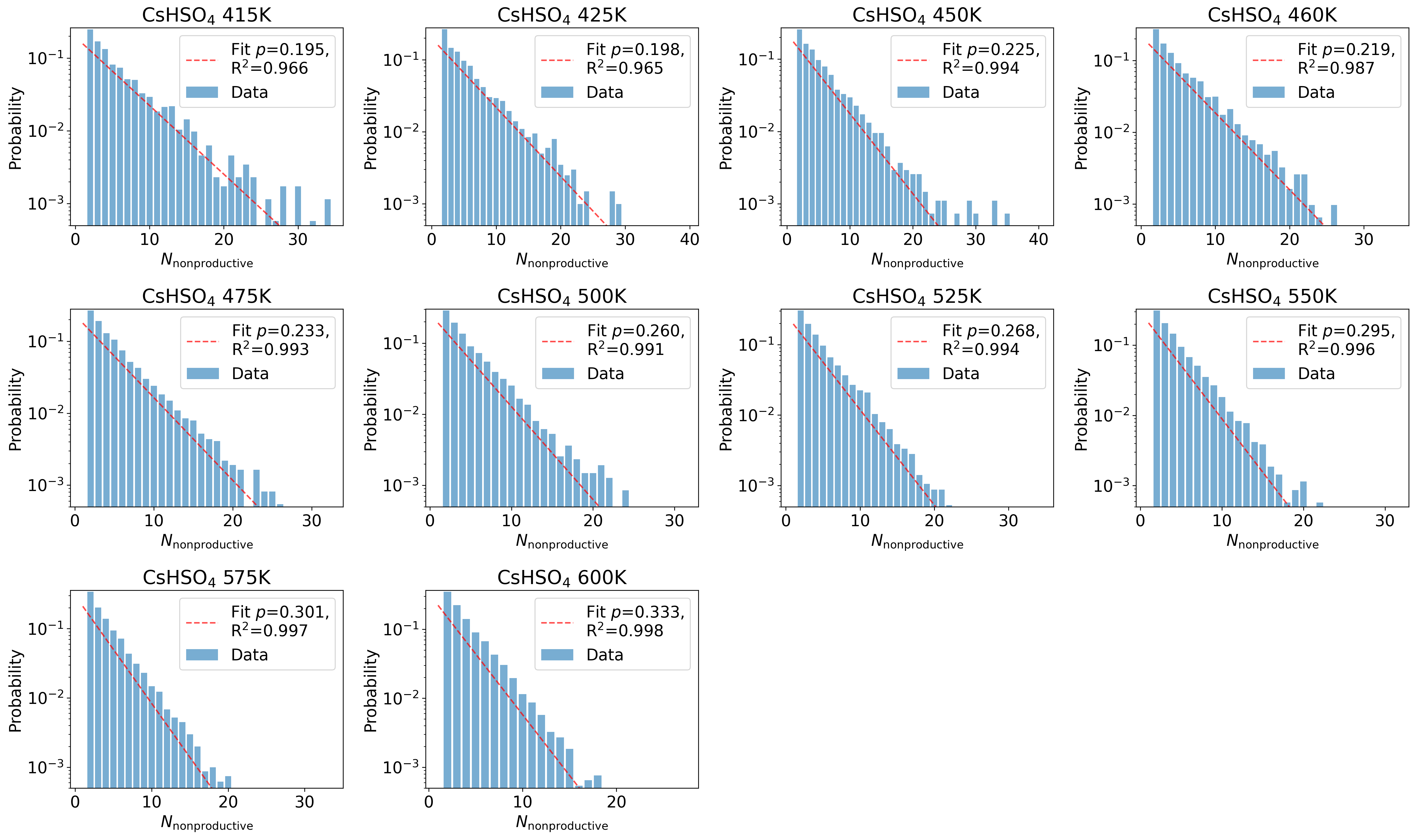}
\caption{Distributions of the number of nonproductive rotations in $\tau_{\mathrm{P-P}}$ intervals from ML MD simulations of \ce{CsH2PO4} (top) and \ce{CsHSO4} (bottom). Dashed lines represent fits to the geometric distribution, which yield the probability of a productive rotation ($p$) listed for each temperature.}
\label{fig::num_non_productive}
\end{figure}

We chose this slope-fitting method over calculating $p$ from the sample mean ($\bar{N}$). While geometric distribution yielded $p = 1/(1+\bar{N})$, the sample mean was highly sensitive to the sparse sampling of the long-tail (high $N$) events. Given the finite nature of MD simulations, the tail of the distribution was subject to significant statistical noise. To obtain a robust estimate of the underlying probability $p$, we extracted the decay rate from the well-sampled, high-probability (low $N$) regime. Specifically, the linear regression was performed only on values of $N$ where the observed frequency count exceeded 10 events. The results of the fitting are shown in Figure \ref{fig::num_non_productive}.

\newpage
\section{Validation of rotation rates via direct counting} \label{si-sec:global_average}

To verify the robustness of the rates fitted from residence time distributions and the geometric probability model, we performed a consistency check using a model-free direct counting method. In this approach, the global average rotation rate $\bar{\lambda}$ was calculated simply as the total number of observed rotation events $N_{\mathrm{events}}$ normalized by the total simulation time $t_{\mathrm{sim}}$ and the number of protons $N_{\mathrm{H}}$:
\[
\bar{\lambda} = \frac{N_{\mathrm{events}}}{t_{\mathrm{sim}} \times N_{\mathrm{H}}}
\]

Here, $N_{\mathrm{events}}$ represents the aggregate sum of events identified across all individual proton trajectories. Applying the classification criteria from Figure 3, we counted an event as nonproductive ($N_{\text{NP}}$) if the rotation resulted in no change to the proton's associated polyanion pair. Conversely, we counted an event as productive ($N_{\text{P}}$) if the proton migrated to a new polyanion pair. The total counts were the sums of these specific occurrences over all protons in the system.

Quantitatively, the mean absolute percentage difference between the global averages and the model-derived rates ranged from approximately 28\% to 56\% (specifically, 28\% and 41\% for NP and P in \ce{CsH2PO4}; 29\% and 56\% in \ce{CsHSO4}). While this represented a systematic offset, such deviations were reasonable and expected. The direct counting method included every rare, long-residence event in the heavy tail of the distribution, which systematically pulled the average rate lower compared to an exponential fit that captured the characteristic decay. Crucially, the rates remained well within the same order of magnitude and exhibited nearly identical activation energies, confirming that both methods described the same underlying physical processes.

\begin{figure}[htbp!]
\centering
\includegraphics[width=\linewidth]{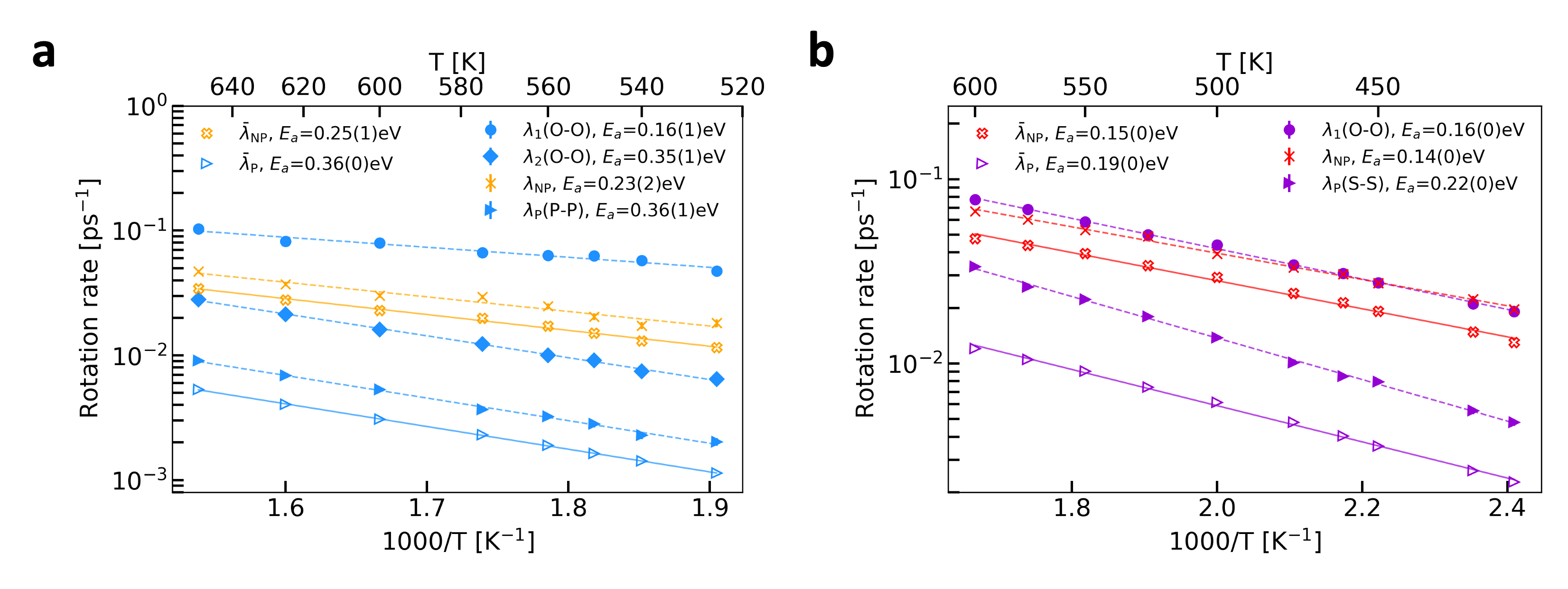}
\caption{Arrhenius plot comparing global average rotation rates (open symbols) with model-derived rates (filled symbols) for (a) \ce{CsH2PO4} and (b) \ce{CsHSO4}. $\bar{\lambda}_{\text{NP}}$ and $\bar{\lambda}_{\text{P}}$ represent the global average nonproductive and productive rates, respectively, calculated via direct counting. Corresponding activation energies are shown for the global average data. Error bars on the filled symbols indicate the standard error from linear regression fits.}
\label{fig::global_average}
\end{figure}

\newpage
\section{Long-range hopping distance distribution}

\begin{figure}[htbp!]
\centering
\includegraphics[width=\linewidth]{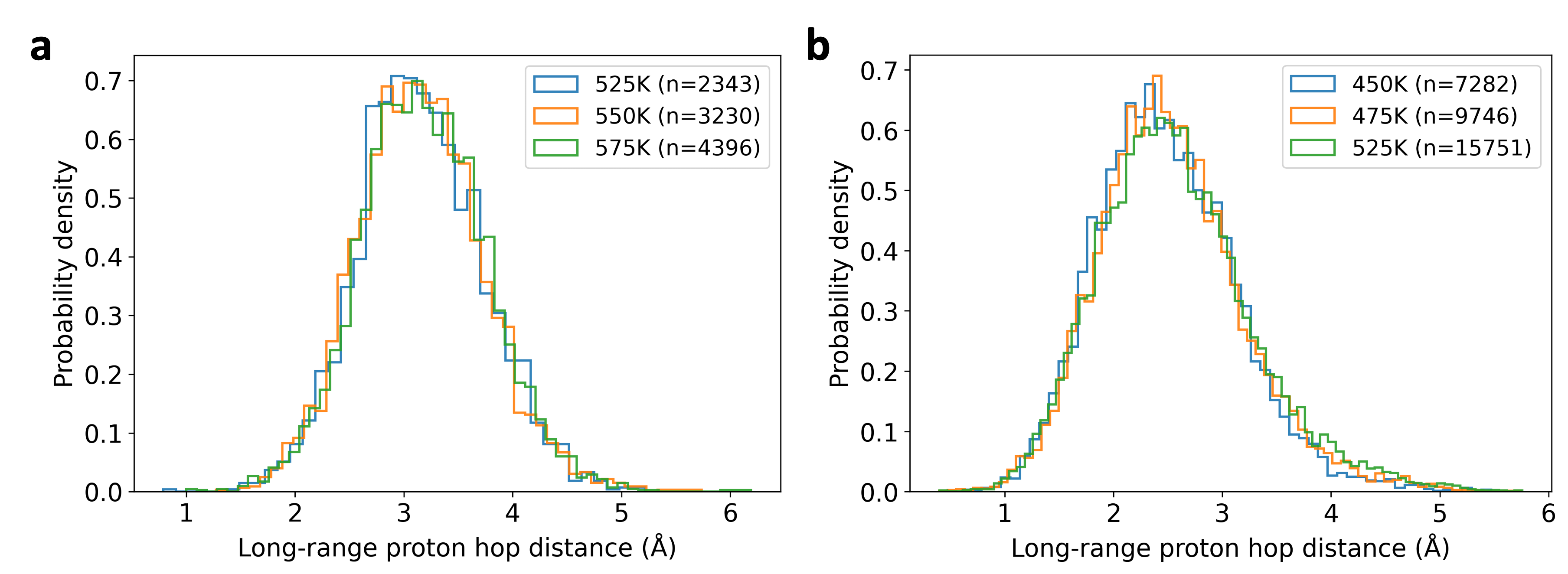}
\caption{Distribution of long-range proton hopping distances during the productive polyanion rotation periods in (a) \ce{CsH2PO4} and (b) \ce{CsHSO4}. The proton hopping distance is calculated as the displacement of the proton from the onset of a productive rotation ($t_s$) to the end of the rotation ($t_f$). The average hopping distance is 3.16 Å at 525 K for \ce{CsH2PO4} and 2.49 Å at 450 K for \ce{CsHSO4}. The temperatures of ML MD simulations and the number of proton jumps $(n)$ are indicated in the labels. }
\label{fig::jump_dist}
\end{figure}

Unlike Figure 3c, which tracks cumulative displacement relative to the simulation start ($t=0$) to identify hopping events, this figure quantifies the net transport of individual hops by measuring displacement strictly over the rotation duration $[t_s, t_f]$.

\newpage
\section{dH/dO ratios over a rotation period}

\begin{figure}[htbp!]
\centering
\includegraphics[width=0.8\linewidth]{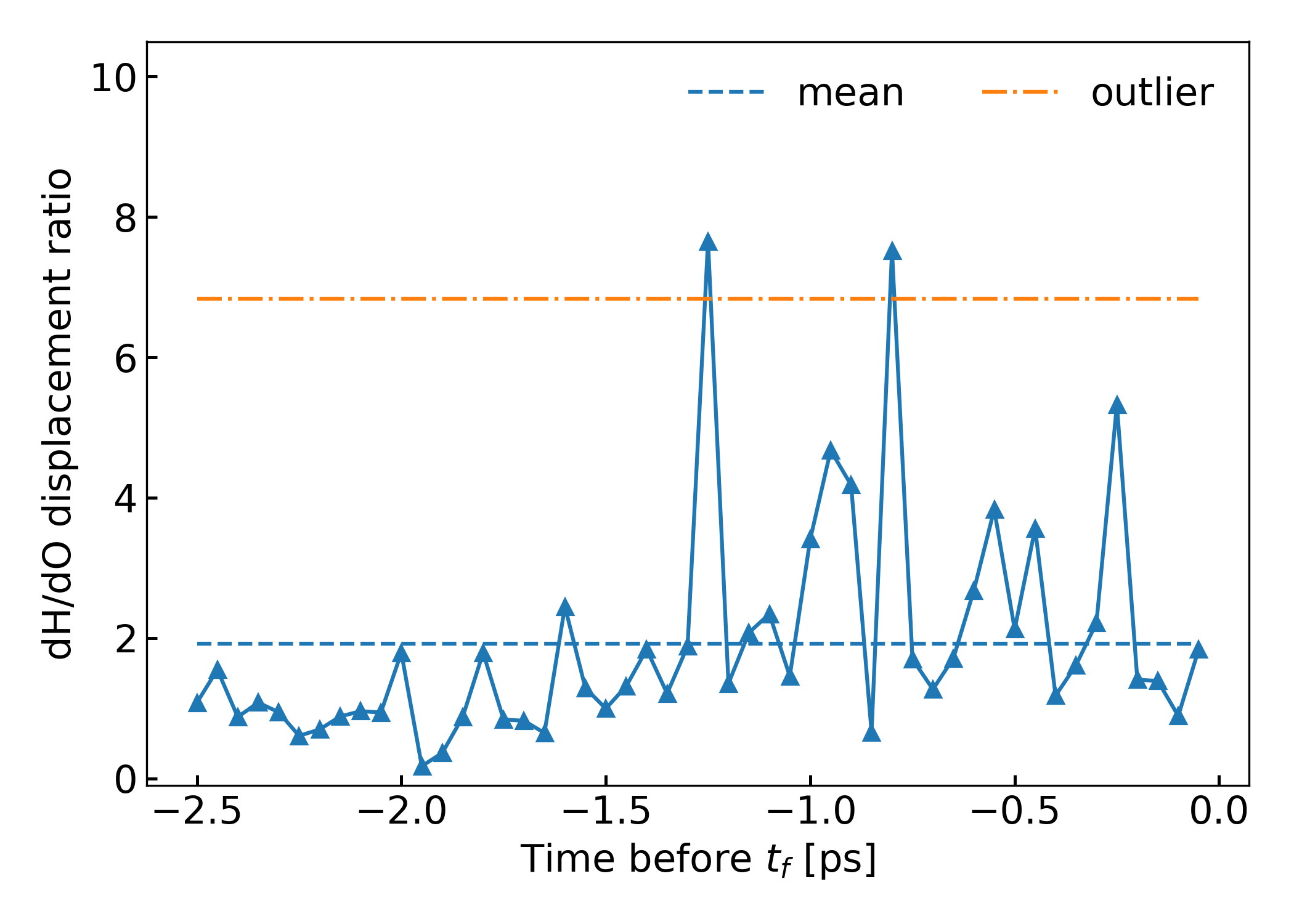}
\caption{Evolution of dH/dO ratios, defined as $\Delta r_{\text{H}}(t_n)/\Delta r_{\text{O}}(t_n)$ for $t_n \in [t_s,t_f]$, tracked for a single proton during its polyanion rotation period. The blue dashed line represents the baseline \textbf{mean} ratio over this rotation period. The orange dot-dashed line indicates the \textbf{outlier} metric, calculated as the average of all ratios with a $Z$-score greater than 2. This outlier value serves as a quantitative indicator of significant \ce{O-H} bond reorientation events.}
\label{fig::OHR_over_time}
\end{figure}

Figure \ref{fig::OHR_over_time} shows the time evolution of the dH/dO displacement ratio for a single proton during its carrier/host polyanion rotation periods. To quantify the intensity of proton motion relative to the oxygen host, we extracted two key metrics for each period: (1) the \textbf{mean} ratio, which established a baseline for coupled motion, and (2) an \textbf{outlier} value (calculated from ratios with a $Z$-score $>2$), which captured the magnitude of significant \ce{O-H} bond reorientations.

This displacement-ratio approach offered a distinct advantage over standard angular difference methods. A simple measurement of the angular change in the \ce{O-H} bonds between consecutive MD snapshots could yield false positives by conflating rigid-body polyanion rotation with actual bond reorientation. For instance, a rotating \ce{P-O} bond with a fixed \ce{P-O-H} angle would show an angular difference in \ce{O-H} bond orientation between MD snapshots, despite no actual \ce{O-H} reorientation occurring. By normalizing the proton's displacement by the oxygen's displacement, our method filtered out these rigid-body contributions, ensuring that we detected only reorientations of the \ce{O-H} bond.

\newpage
\section{\texorpdfstring{\ce{O-H}}{O-H} bond reorientation}

\begin{figure}[htbp!]
\centering
\includegraphics[width=0.8\linewidth]{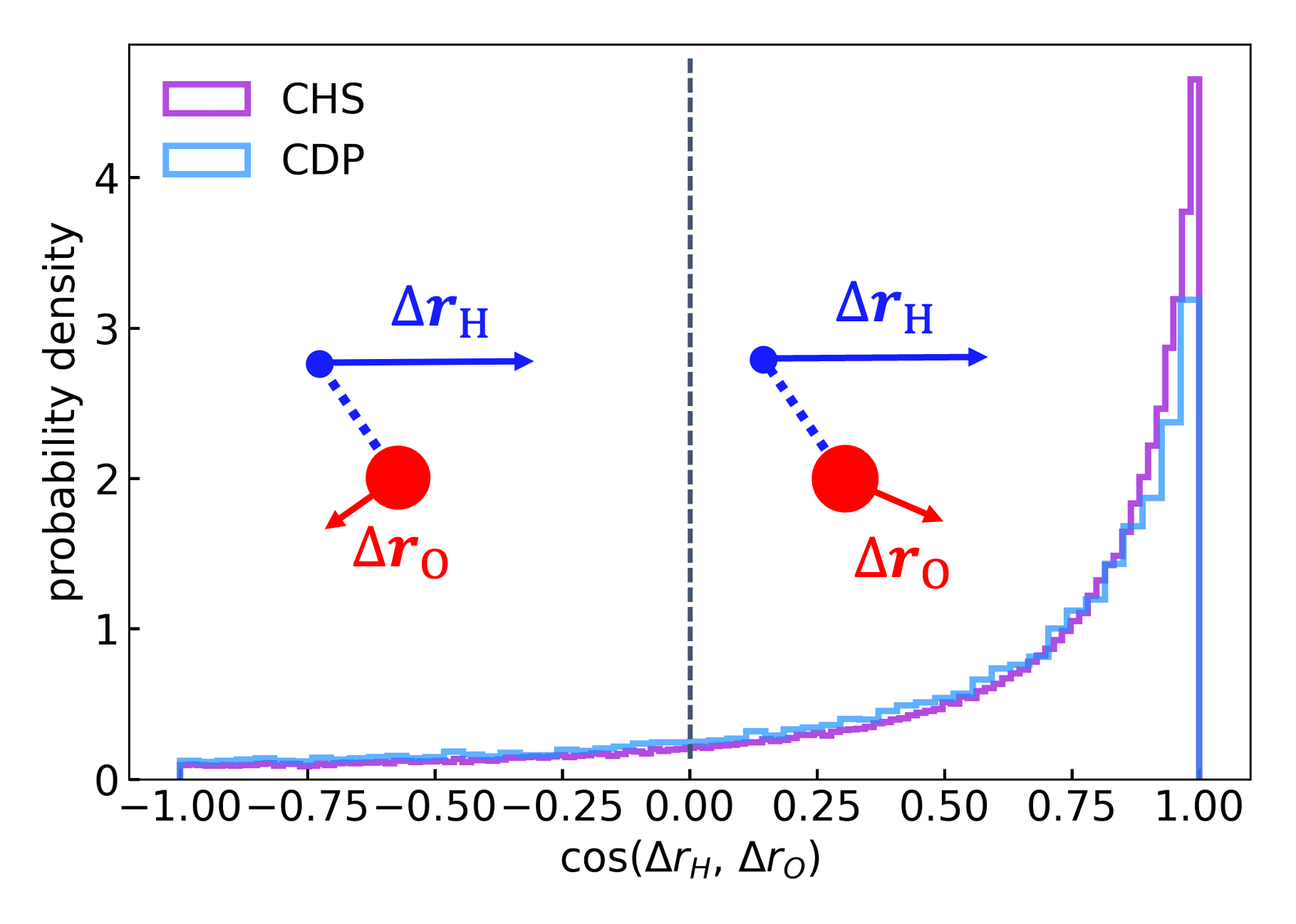}
\caption{The distribution of cosine angles between the instantaneous H displacement and the instantaneous O displacement, defined as $\cos (\Delta \vb*{r}_{\text{H}}(t_n), \Delta \vb*{r}_{\text{O}}(t_n))$, during the polyanion rotation period.}
\label{fig::cos_HO_CHS}
\end{figure}

Figure \ref{fig::cos_HO_CHS} demonstrates that in both \ce{CsHSO4} and \ce{CsH2PO4}, the hopping proton predominantly moves in the direction of carrier rotation.

\newpage
\section{Proton trajectory of \texorpdfstring{\ce{CsHSO4}}{CsHSO4}}

\begin{figure}[htbp!]
\centering
\includegraphics[width=0.8\linewidth]{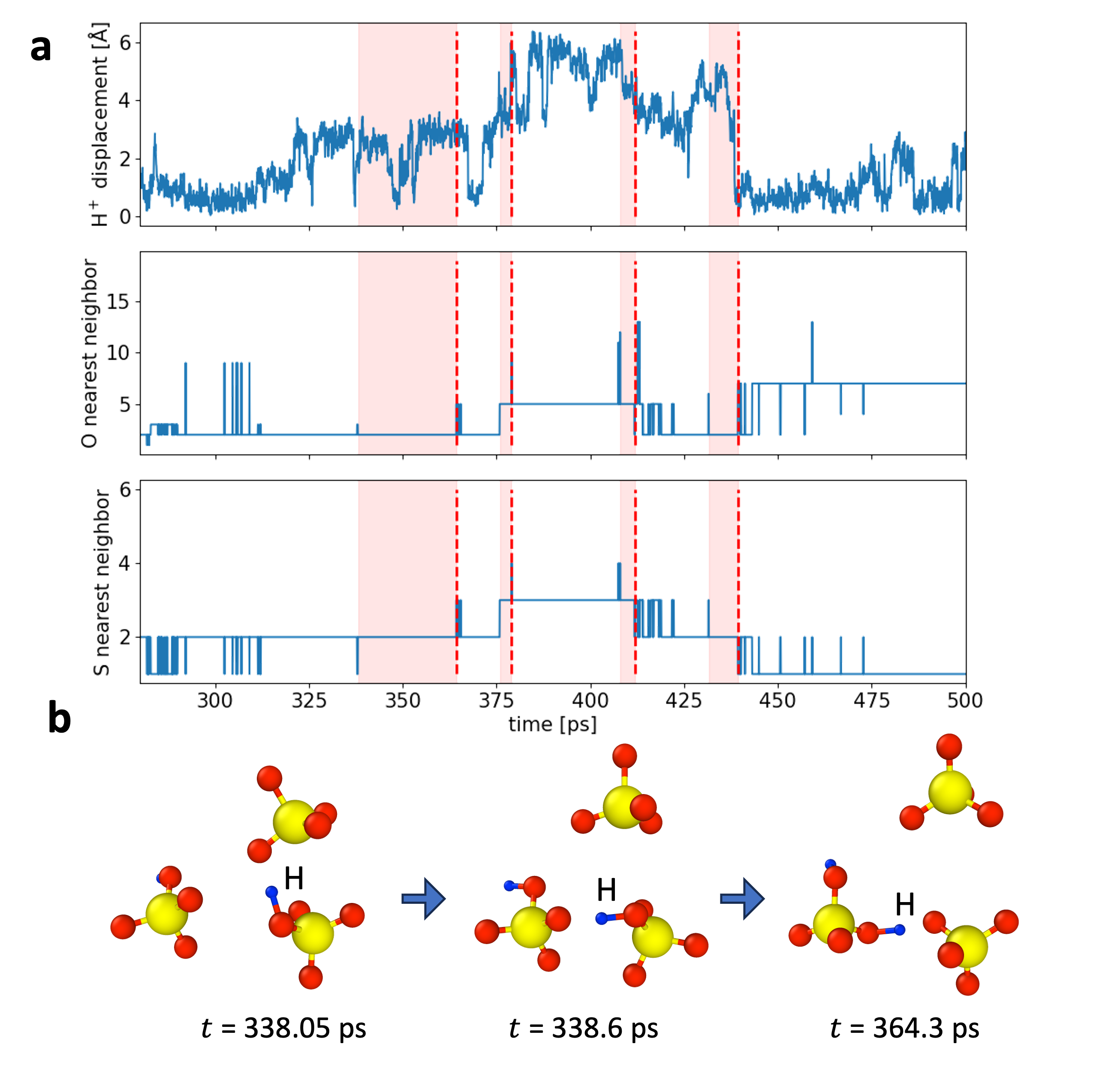}
\caption{(a) Time evolution of proton dynamics from an ML MD simulation of \ce{CsHSO4} at 500 K showing (from top to bottom): the proton displacement relative to $t=0$, its nearest \ce{O} anion neighbors, and its nearest \ce{SO4} polyanion neighbors. Shaded regions indicate rotation periods where the covalent \ce{O-H} bond is maintained. (b) Snapshots of the hopping proton at 338.05, 338.6, and 364.3 ps. Proton orients toward the next polyanion 25.7 ps before hopping. }
\label{fig::proton_trj_CHS}
\end{figure}

Figure \ref{fig::proton_trj_CHS} shows that protons in \ce{CsHSO4} tend to bond more persistently with a single O anion, resulting in a slower rattling rate compared to \ce{CsH2PO4}. The red dashed lines indicate S-pair switch events, which occur following proton displacement jumps, as this involves the additional step of breaking the \ce{O-H} bond.

\newpage
\section{Polyanion rotation distances in \texorpdfstring{\ce{CsHSO4}}{CsHSO4}}

\begin{figure}[htbp!]
\centering
\includegraphics[width=0.8\linewidth]{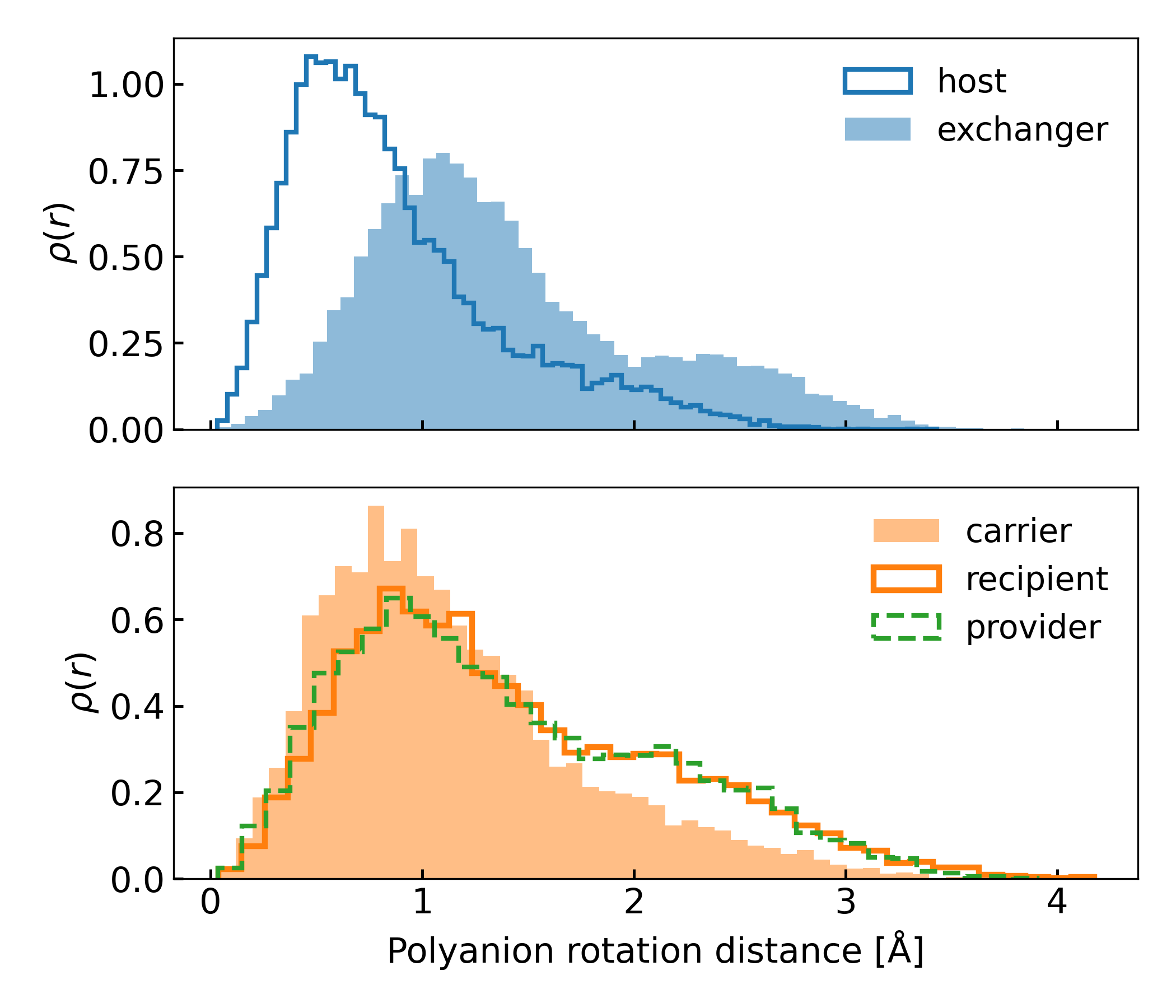}
\caption{Comparison of rotational magnitudes between nonproductive (top) and productive (bottom) rotation mechanisms in a 450 K \ce{CsHSO4} simulation. Shaded histograms represent the rotating polyanions (``exchanger" and ``carrier"). In nonproductive events, the proton-carrying ``host" remains mostly stationary while the ``exchanger" rotates. In productive events, the proton-carrying ``carrier" itself undergoes rotation.}
\label{fig::CHS_rotation}
\end{figure}

\section{\texorpdfstring{\ce{SO4}}{SO4} orientations}

In superprotonic \ce{CsHSO4}, \ce{SO4} tetrahedra alternate orientations along the 
c-axis. We applied symmetry operations to align these tetrahedra to a common reference frame, enabling direct comparison of polyanion orientations at later times.

\clearpage
\renewcommand\refname{Supplementary References}
\bibliographystyle{naturemag}
\bibliography{pconduct.bib}